\documentclass[manuscript,screen, authorversion,nonacm]{acmart}

\usepackage{tcolorbox}
\tcbuselibrary{breakable}
\usepackage{tikz}
\usetikzlibrary{positioning}
\usetikzlibrary{matrix,calc}
\usetikzlibrary{backgrounds}
\usepackage{multirow}
\usepackage{multicol}

\usepackage{colortbl}
\usepackage{enumitem}
\usepackage{ifthen}
\usepackage{tabularx}
\usepackage{tabularray}
\usepackage{rotating}
\usepackage{makecell}
\usepackage{hhline}
\usepackage{nicematrix}
\usepackage{booktabs}
\usepackage{array}
\usepackage{adjustbox}
\usepackage{lipsum}
\usepackage{subfig}
\usepackage{ragged2e} % for \raggedright command
\usepackage{url}

\newcommand{\PreserveBackslash}[1]{\let\temp=\\#1\let\\=\temp}
\newcolumntype{C}[1]{>{\PreserveBackslash\centering}p{#1}}
\newcolumntype{R}[1]{>{\PreserveBackslash\raggedleft}p{#1}}
\newcolumntype{L}[1]{>{\PreserveBackslash\raggedright}p{#1}}

\begin{document}
%\begin{multicols*}{2} 
\title[Exploring conversational AI and response types for information elicitation]{``\textit{If we misunderstand the client, we misspend 100 hours}'': Exploring conversational AI and response types for information elicitation}

%%Authors
\author{Daniel Hove Paludan}
\orcid{0009-0006-3927-9944}
\email{dpalud20@student.aau.dk}
\affiliation{%
  \institution{Aalborg University}
  \city{Aalborg}
  \country{Denmark}
}
\author{Julie Fredsgård}
\orcid{0009-0006-6652-3957}
\email{jfreds20@student.aau.dk}
\affiliation{%
  \institution{Aalborg University}
  \city{Aalborg}
  \country{Denmark}
}
\author{Kasper Patrick Bährentz}
\orcid{0009-0002-9599-5815}
\email{kbahre20@student.aau.dk}
\affiliation{%
  \institution{Aalborg University}
  \city{Aalborg}
  \country{Denmark}
}

\author{Ilhan Aslan}
\email{ilas@cs.aau.dk}
\orcid{0000-0002-4803-1290}
\affiliation{%
  \institution{Aalborg University}
  \city{Aalborg}
  \country{Denmark}
}

\begin{teaserfigure}
    \centering
    \includegraphics[width=\textwidth]{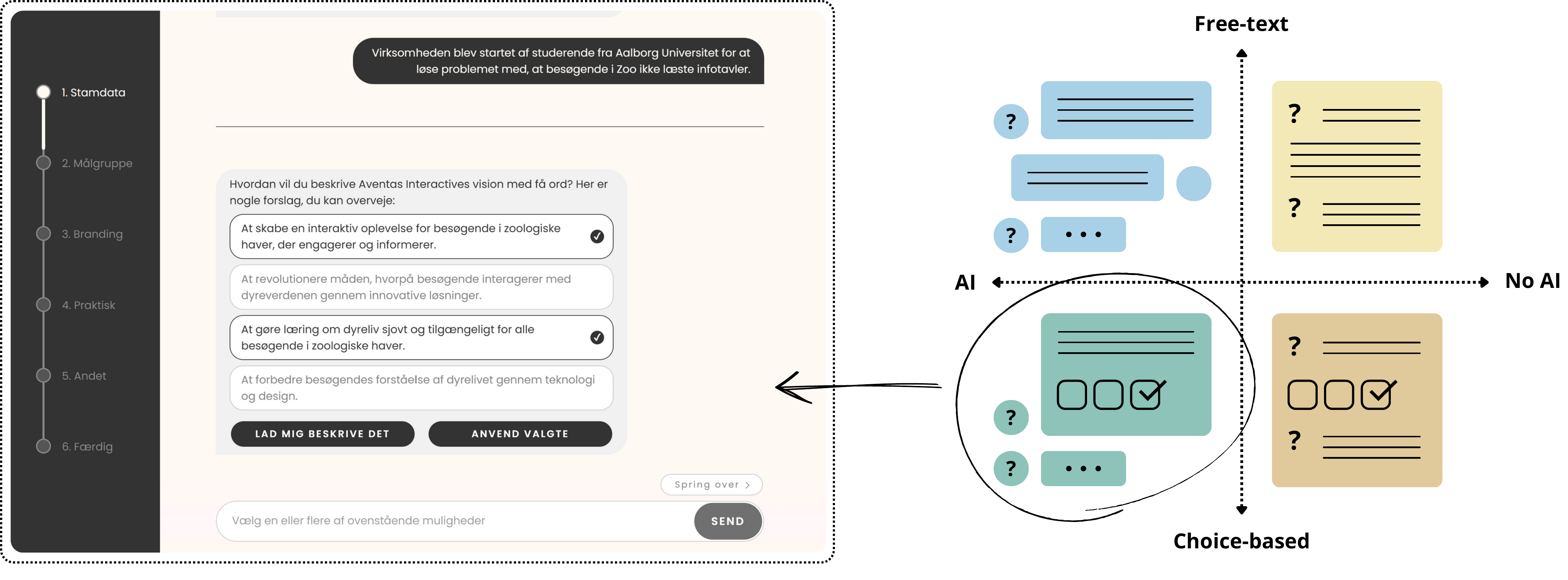}%
    \Description{Teaser figure.}
    \caption{An illustration of the four experimental conditions of the developed elicitation tool to support mutual understanding in early-stage client-designer collaboration (right), highlighting its integration of conversational AI and choice-based responses (left).}
    \label{fig:teaser_figure}
    \vspace{1em}  % adjust as needed
\end{teaserfigure}

\begin{abstract}
Client-designer alignment is crucial to the success of design projects, yet little research has explored how digital technologies might influence this alignment. To address this gap, this paper presents a three-phase study investigating how digital systems can support requirements elicitation in professional design practice. Specifically, it examines how integrating a conversational agent and choice-based response formats into a digital elicitation tool affects early-stage client-designer collaboration. The first phase of the study inquired into the current practices of 10 design companies through semi-structured interviews, informing the system's design. The second phase evaluated the system using a 2x2 factorial design with 50 mock clients, quantifying the effects of conversational AI and response type on user experience and perceived preparedness. In phase three, the system was presented to seven of the original 10 companies to gather reflections on its value, limitations, and potential integration into practice. Findings show that both conversational AI and choice-based responses lead to lower dependability scores on the \textit{User Experience Questionnaire}, yet result in client input with greater clarity. We contribute design implications for integrating conversational AI and choice-based responses into elicitation tools to support mutual understanding in early-stage client-designer collaboration.

\end{abstract}

\keywords{}

%\settopmatter{printfolios=true}

\maketitle

\section{Introduction} \label{Introduction}
%\setlength{\parindent}{8pt}

% ________ State of the world ___________
Effective communication between designers and clients is necessary for a project to be successful. If clients and designers do not agree on the design space of potential solutions, both parties will waste resources on designs that miss the mark \cite{lee2024clientdesignernegotiation, dafaalla2022suitablerequirements}. As each serve different roles in the design process---one creating the product, the other defining its purpose---clients and designers often carry very different mental models \cite{goldschmidt2007mentalmodel}. Yet for a design outcome to satisfy both parties, clients and designers must share their mental models to a sufficient degree, and achieving such sharedness is not trivial \cite{goldschmidt2007mentalmodel, casakin2017sharedness}. To facilitate alignment, design briefs and other artefacts are paramount, enabling stakeholders to externalise their individual views and combine their understandings into a coherent whole \cite{goldschmidt2007mentalmodel, dix2011externalisation}.

While artefacts can aid in reaching a shared understanding, the diversity of clients and variability between projects make it difficult to apply standardised approaches \cite{lee2024clientdesignernegotiation}. Design briefs, in particular, demand significant time and effort, and are often hindered by poor information transmission and unclear, or unresolved, client requirements \cite{zhu2024aidesignbriefcreation}. Seeking to accommodate such limitations, vast amounts of literature has explored how artificial intelligence (AI) might be leveraged to support designers in their work, the majority focusing on facilitating divergent and convergent thinking in later stages of the design process \cite{lee2024when}. For example, in previous research \cite{paludan2025} we studied prompt-based image generation AI from a ``more than human'' \cite{Giaccardi04072025} design perspective, aiming to understand the relation and dilemma  between energy usage, and creativity support. While the usage of AI in design processes is not straightforward, on the whole, these studies reveal the potential of AI in supporting designers' creative processes---facilitating both idea generation, decision-making, and evaluation \cite{lee2024when, tang2025preandpostai}.

For a creative process to be meaningful, it is crucial for practitioners to have the right starting point. Because design work is only valuable insofar as it solves the right problem, clients and designers must agree on both the \textit{what} and the \textit{why} of a project \cite{lee2024clientdesignernegotiation}. If the wrong trajectory is set from the outset, design projects are headed for failure \cite{dafaalla2022suitablerequirements}. To prevent this, the involved parties must adapt, adjust, or even completely revise their mental models in order to achieve alignment \cite{casakin2017sharedness}. In this process, communication is key, shaping not only the design process but also its outcome. Specifically, it is important that both clients and designers are able to effectively coordinate the contents of their individual mental models \cite{casakin2017sharedness}. While AI has proven capable of supporting requirements elicitation \cite{jacobsen2025chatbots, ataei2024elicitron, kim2019comparing}, its potential in facilitating communication between designers and clients remain largely unexplored.

To explore how AI might support such communication, we conducted semi-structured interviews with 11 experienced designers from 10 small design companies, identifying client-designer alignment as an essential aspect of project success. Building on these insights, we set out to investigate how information elicitation systems can be designed to support clients and designers in developing a shared understanding of their project space. In doing so, we sought to answer the following research questions:

\begin{itemize}
    \item [\textbf{RQ1}] \textit{How can a system be designed to support information elicitation in early-stage client-designer collaboration?}
    \item [\textbf{RQ2}] \textit{How does the integration of conversational AI and choice-based response formats into an information elicitation tool influence clients' user experience and perceived preparedness?}
\end{itemize}

To investigate these research questions, we developed four variants of an information elicitation tool for clients to use before their initial meeting with a designer. These variants were evaluated in a 2x2 between-subjects design, examining the effect of \textit{response type} (free-text vs. choice-based) and \textit{AI presence} (AI vs. no AI) on clients' user experience and perceived preparedness.

50 participants took part in the evaluation, using the system to elicit information on hypothetical design projects. The quality of the provided information was assessed through the lens of Gricean Maxims\footnote{As adapted from Xiao et al. \cite{xiao2020chatbotSurvey} and Jacobsen et al. \cite{jacobsen2025chatbots}}, while user experience (UX) was measured via the \textit{User Experience Questionnaire} (UEQ). Semi-structured interviews were conducted with seven of the ten design companies consulted during the system's design phase, providing qualitative feedback on its usability and applicability in practice.

We found that both AI presence and choice-based responses improved the clarity of client responses, though they resulted in significantly lower dependability scores on the UEQ---and, in case of AI presence, also in significantly lower perceptions of efficiency. Qualitative insights suggest that elicitation tools can benefit from embedded AI assistants when integrated thoughtfully; adapting to both clients and designers while respecting designer expertise.

Based on these insights, we propose the following design implications for information elicitation tools in early-stage client-designer collaboration:

\begin{enumerate}
    \item Design for confident information sharing through client control and role-aligned input
    \item Accommodate diverse preferences through customisation for both clients and designers
    \item Support layered and editable AI outputs that synthesise client responses
\end{enumerate}

\section{Related work} \label{sec:related_work}

\subsection{Facilitating alignment through design briefs}
Lee et al. describe communication between clients and designers as the challenge of aligning two different mental models of the envisioned project space \cite{lee2024clientdesignernegotiation}. While it is the job of designers to create a satisfying product, it is the job of clients to provide information on requirements, opinions, and preferences \cite{casakin2017sharedness}. Clients and designers must, therefore, adapt, adjust, and negotiate any aspect of a project on which their individual mental models are not sufficiently aligned (e.g., its goals, scope, or budget) \cite{casakin2017sharedness,lee2024clientdesignernegotiation}.

Often, this negotiation involves the creation of design briefs and other artefacts (e.g., sketches \cite{goldschmidt2007mentalmodel}), setting the expectations for the project \cite{lee2024clientdesignernegotiation}. Such artefacts can be crucial, as inadequate requirements elicitation---i.e., failing to help clients figure out what they want---can have significant consequences, ranging from design misalignment \cite{ataei2024elicitron} to increased development costs \cite{murugesan2017elicitation} and, ultimately, project failure \cite{dafaalla2022suitablerequirements}. Design briefs define what is to be built---but not how to do it---serving as a form of contract between client and designer \cite{koronis2019designbrief}. They are a critical tool for organising the resources and requirements of a design project; taking shape in the initial stages of a project as a synthesis of a client's preliminary ideas and the expert knowledge of the designer \cite{zhu2024aidesignbriefcreation}.

Koronis et al. examined the factors that constitute a good design brief and found that this depends on its purpose: if the goal is to promote novelty, visual stimuli of existing solutions should be avoided; if the goal is appropriateness, the use of quantitative product requirements (e.g., price range, size) can be very effective \cite{koronis2021crafting}. To align on a design's physical properties, Goldschmidt argues that product visualisations are indispensable, enabling team members to share knowledge, reason about it, and combine it into a coherent whole---in short, making it possible for them to see eye to eye \cite{goldschmidt2007mentalmodel}. While visualisations can thus facilitate a shared understanding, they are not neutral; rather, their form plays a significant role in shaping the structure of a team's mental model \cite{schnotz2008externalrepresentations}. More broadly, the form and content of inspiration sources have been shown to influence design outcomes to varying degrees, depending on their modality, level of detail, and distance to the design problem \cite{alipour2018fixation}. These influences stem not only from the source itself, but also from the problem (e.g., its formulation), the designer (e.g., their level of expertise), and the design process (e.g., the employed model of thought), making it difficult to establish unified guiding principles.

In the construction of design briefs, Lee et al. note how client goals may be insufficiently detailed, requiring designers to probe for deeper information \cite{lee2024clientdesignernegotiation}. As the questions asked by designers can steer a project's direction, they suggest practitioners to deliberately control what information they elicit from clients---for example, asking about preferred mood and tone rather than examples of preferred aesthetics---so as to maintain creative freedom to benefit both their clients and themselves \cite{lee2024clientdesignernegotiation}. Zhu et al. found that designers have two primary hurdles formulating design briefs: one involves interpreting background knowledge (e.g, unfamiliar project domains and technical gaps) and the other involves interpreting human factors (e.g., communicating with complex stakeholders) \cite{zhu2024aidesignbriefcreation}.

While design briefs can reduce redundant work and save costs, they are a substantial investment in terms of time and effort and are challenged by poor information transmission, changing client requests, and cost pressures. To investigate how AI might help address these challenges, Zhu et al. engaged senior UX design experts in generating design briefs with the assistance of ChatGPT. The results demonstrated a significant reduction in task completion time and an enhanced user experience for the designers \cite{zhu2024aidesignbriefcreation}. While using large language models (LLMs) comes with downsides, such as hallucinations \cite{yoon2024can}, the results illustrate the potential value of employing AI in designers' daily work.

\subsection{AI tools in design practice}
From supporting data-driven decision-making and generating prototypes, to anticipating customer needs and facilitating personalised interactions, the integration of AI into design practice promises both efficiency and creativity \cite{razmerita2022aicollaboration, uusitalo2024clay, lee2024when}. Yet successful adoption demands entirely new skill sets, requiring designers to relearn some of the profession's core competencies \cite{uusitalo2024clay}. Specifically, designers looking to employ generative AI in their practice need to sharpen their verbal creativity skills for effective prompt creation \cite{tan2024using, akverdi2024generative, lee2024impactsketchprompt}.

Through case studies of industrial designers, Tang et al. found that human creativity and intuition are essential in both divergent and convergent phases of design, emphasising the importance of incorporating AI at the right time and in the right way \cite{tan2024using}. Others similarly recognise the complementary strengths of humans and AI, calling for cooperative interaction through human-AI collaboration \cite{memmert2022humanaicollaboration}. Zhou et al. argue that human-AI collaboration can mitigate some of the shortcomings of human-human collaboration \cite{zhou2024understanding}---e.g., by enabling real-time visual outcomes, as demonstrated by Tan and Luhrs \cite{tan2024using}. They call for AI systems to foster in-depth discussions of design requirements, encourage users to double-check possible misunderstandings, and offer alternative solutions as sources of inspiration \cite{zhou2024understanding}. Lee et al. claim that future AI-design support systems need to consider a more interactive involvement of AI, taking a human-centric approach \cite{lee2024when}. This is a view supported by Uusitalo et al., who highlight the human ability to perceive, understand, and respond to user needs, preferences, and emotions, as well as the broader context for which a product is designed \cite{uusitalo2024clay}. We have argued in previous work \cite{paludan2025} to consider the impact on energy usage when designing interfaces for designers, which include generative AI features.  

While AI can enhance efficiency by automating time-consuming tasks, Akverdi and Baykal raise concerns regarding its potential to constrain creativity within algorithmic patterns; their study indicates that the rapid output of AI tools may discourage instinctual exploration and independent ideation \cite{akverdi2024generative}. On the contrary, Chiou et al. found that the incorporation of AI into design processes opens up new avenues for artistic expression, as AI input contributes a fresh form of self-expression \cite{chiou2023designing}. On the whole, the successful integration of AI tools can provide designers more time for what they do best, but it must be approached thoughtfully, requiring designers to adapt, develop new skill sets, and take control when necessary.

\begin{figure}[ht!]
    \centering
    \includegraphics[width=0.9\linewidth]{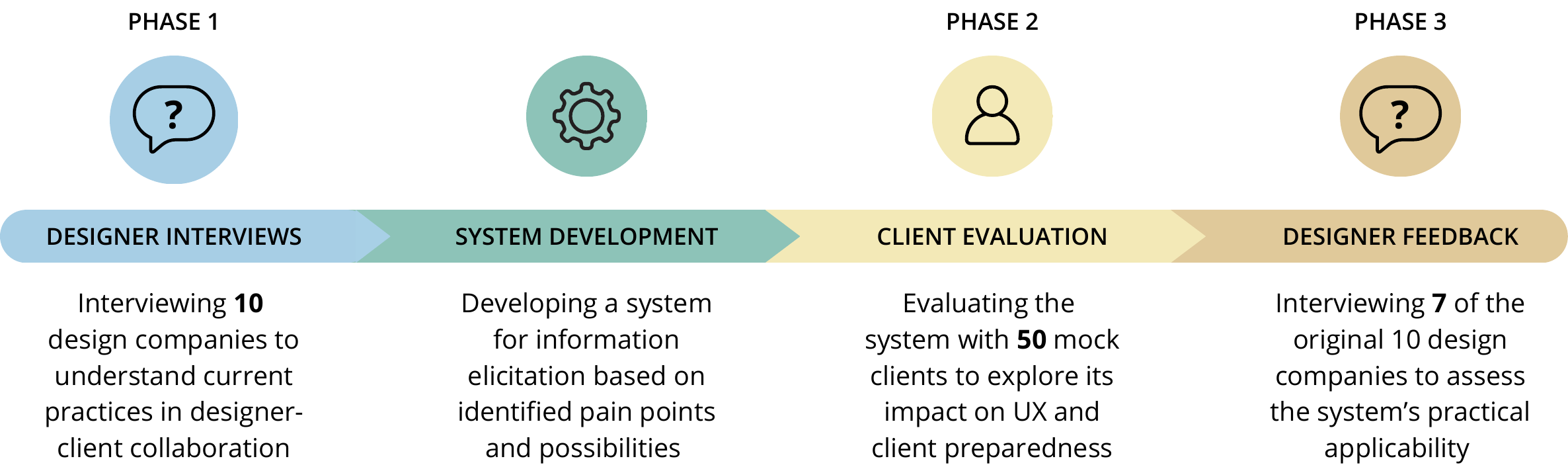}
    \caption{The three phases of the conducted study.}
    \label{fig:process}
\end{figure}

\subsection{Virtual agents in requirements elicitation}
Razmerita et al. highlight the potential in using conversational AI---systems designed to act or react as humans in a dialogue session---to coordinate, communicate, and broker knowledge \cite{razmerita2022aicollaboration}. Much research suggests that these qualities can enhance requirements elicitation, accommodating some of the needs that traditional methods are unable to \cite{ataei2024elicitron, kim2019comparing, xiao2020chatbotSurvey}.

Traditional methods for requirements elicitation face a trade-off between depth and scale: while interviews can yield rich, detailed insights, they are significantly more difficult to conduct at scale compared to structured surveys \cite{wuttke2024transformingsurveys}. Web surveys, by contrast, are inherently limited compared to face-to-face conversations, as they lack the ability to pose dynamic follow-up questions, clarify ambiguities, or assess the sincerity of responses \cite{kim2019comparing}. To address this trade-off, Wuttke et al. explored the potential of replacing human interviewers with LLMs, finding that LLM-based interviewing is a viable approach to producing quality data at scale \cite{wuttke2024transformingsurveys}. They emphasise the importance of tailoring interviewer behaviour to the specific research context---e.g., adjusting the frequency of follow-up questions or the level of formality. Complementary research demonstrates the ability of conversational AI to support requirements elicitation, uncover latent needs \cite{ataei2024elicitron}, and prepare interviewers by providing them with information outside their area of expertise \cite{lauer2024vectorpersonas, liu2025interview}.

The limitations of traditional web surveys have led some researchers to examine how AI might make surveys more dynamic, personalised, and effective \cite{jacobsen2025chatbots, kim2019comparing, rhim2022surveychatbots}. Krosnick proposed that survey respondents tend to provide satisfying responses---i.e., responses that meet minimal acceptability standards---rather than accurate ones, as delivering precise and sincere responses demands significant cognitive effort \cite{krosnick1991responsesatisfying}. Such satisfying behaviour has been found to be especially prevalent in web surveys \cite{heerwegh2008responsesdifferences}, and Kim et al. hypothesise that this stems from their lack of interactivity \cite{kim2019comparing}. They found that the conversational interactivity added by virtual agents produces higher quality data by decreasing respondents' satisficing behaviour \cite{kim2019comparing}. These conclusions are supported by emerging research on survey-embedded chatbots, suggesting that the asking of dynamic follow-up questions can reveal deeper insights into participants' experiences, potentially surpassing traditional open-ended questions \cite{jacobsen2025chatbots, hu2024designing}. 

While integrating conversational AI into web surveys is an increasingly popular approach, users generally anticipate lower communication quality when interacting with chatbots, perceiving them as less sociable, empathetic, and interactive than humans \cite{zhou2023talkingbot}. The majority of people prefer to communicate with people, as these better understand them and are therefore better able to provide relevant answers \cite{ashfaq2020chatbotsatisfaction}. The usage of elicitation techniques to design novel interaction designs \cite{Wobbrock},  by considering different use contexts and design choices \cite{Aslan2018}, as well as role-playing \cite{Bittner2019} are established design and research practice. However, these practices are guided and observed by human researchers or designers. Since users respond and focus differently when interacting with a chatbot, as opposed to a human, multiple authors highlight the advantage of incorporating anthropomorphic characteristics into chatbots to facilitate communication that mimics interpersonal conversations \cite{zhou2023talkingbot, rhim2022surveychatbots}. This can lead to higher levels of self-disclosure, but may also increase social desirability bias---i.e., the degree to which respondents' change their answer to conform to social norms.

Overall, AI tools show promise as a means of supporting practitioners in both requirements elicitation, ideation, and design, with virtual agents in particular enabling an unprecedented bridging of quantitive and qualitative data collection. While extensive research underscores the importance of aligning the mental models of clients and designers, the role of AI in facilitating this alignment remains underexplored. To address this gap, this paper examines how elicitation tools can be designed to facilitate client-designer alignment in the early stages of project collaboration.

\section{Phase 1: Understanding the practices of client-designer communication} \label{sec:pilot}
Building on insights from related work, we conducted an exploratory study to better understand how alignment is pursued in practice between design practitioners and their clients. These interviews comprised the first of three study phases, as can be seen in Figure \ref{fig:process}. Through semi-structured interviews, this phase aimed to uncover pain points in current practitioner workflows and, using inspiration cards (as inspired by Nadia Piet \cite{piet2020aicards}), engaged participants in envisioning how AI might support or alleviate these challenges.

\subsection{Participants}
Interviews were conducted with 11 professionals (10 male, 1 female) from 10 design companies of various sizes (ranging from 1 to 20+ employees), including industrial designers, graphic designers, and architects. Their professional experience ranged from 10 to 30 years. Recruitment was conducted via email and phone outreach. To be eligible for participation, designers had to use or consider using AI in their design work.

\subsection{Apparatus and procedure}
\begin{figure}[!ht]
\centering
\includegraphics[width=0.5\linewidth]{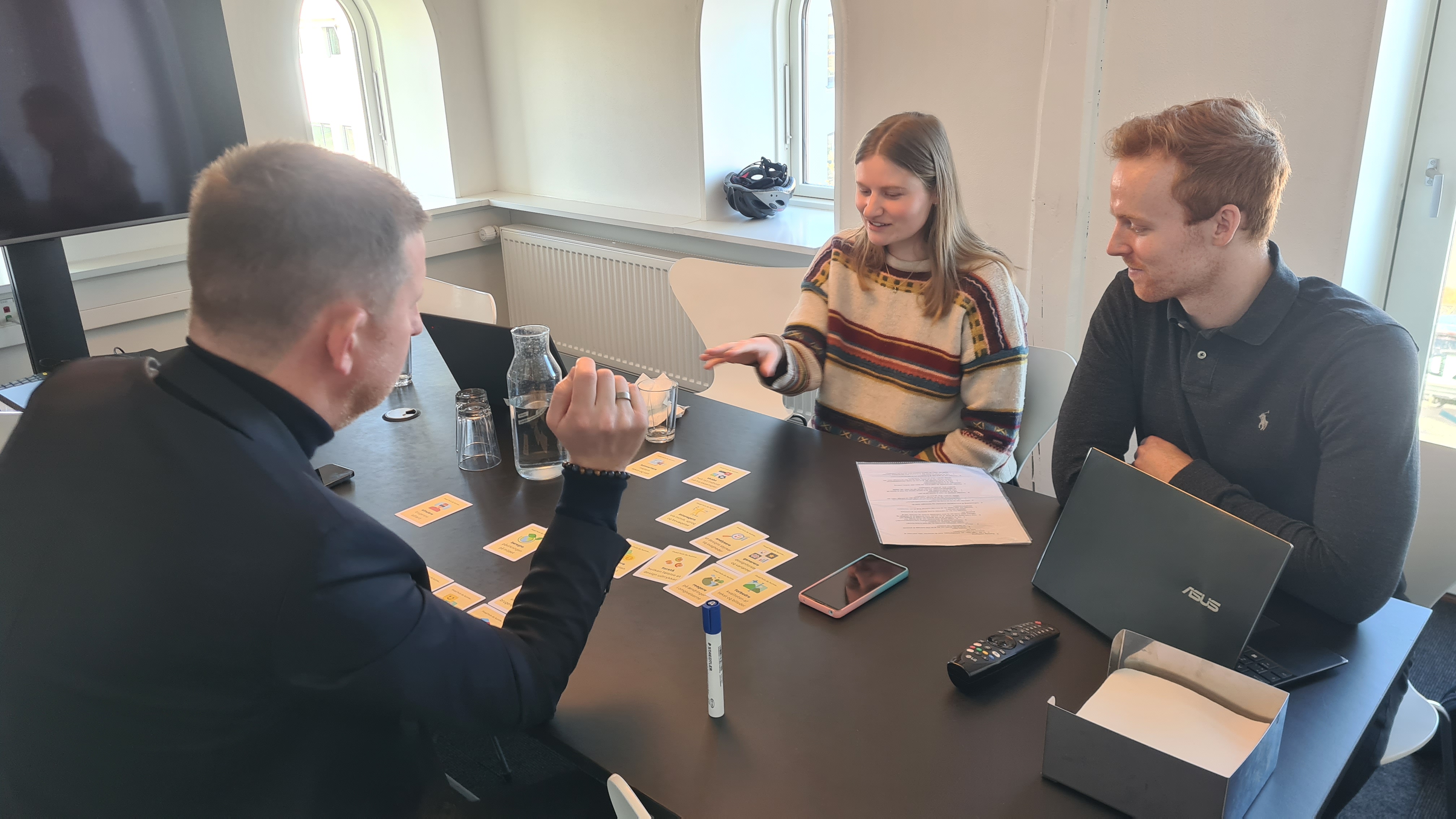}
\caption{The study setup of the phase one designer interviews. One researcher facilitated the interview while another took notes. 18 inspiration cards were arranged on the table, allowing participants to freely select and engage with them.}
\label{fig:study1}
\end{figure}

The interviews were semi-structured and lasted 45 minutes to an hour, with the first half covering practitioners' background, design practices, and client collaboration. The second half focused on the role of AI in participants' current and future practices and introduced inspiration cards to help participants recall, envision, and ideate---particularly in the context of client collaboration. The study setup can be seen in Figure \ref{fig:study1}.

\begin{figure}[!ht]
\includegraphics[width=0.6\linewidth]{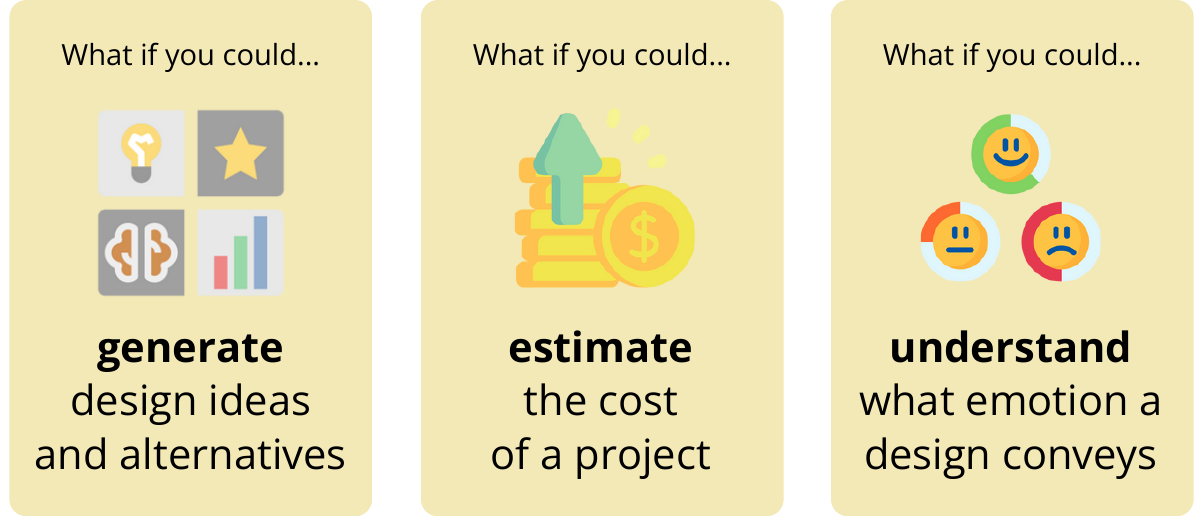}
\caption{Three of the 18 inspiration cards created for exploring how AI might be integrated into designers' current practices of client collaboration.}
\label{fig:inspirationcards}
\end{figure}

The 18 inspiration cards were developed using Piet's \textit{AI Ideation Cards} \cite{piet2020aicards} as a foundation. As can be seen in Figure \ref{fig:inspirationcards}, all cards included the phrase ``\textit{What if you could...}'', followed by a statement highlighting one potential use of AI. Some of these were specific while others were abstract, leaving room for participants to interpret them in ways suiting their specific practice.

Seven interviews were conducted on-site while three were conducted through \textit{Microsoft Teams}. To fine-tune the formulation and order of questions, a pilot interview was conducted with a student of interaction design. In addition to the notes taken during the conversation, interviews were recorded and subsequently transcribed. Participants were also encouraged to share the tools and templates they use in their practice, specifically those concerning client collaboration and those integrating AI. Half the participants provided such artefacts, including design brief templates, brand cards, (custom) AI tools, documents detailing AI experiments, and other documents for aligning client and designer expectations.

\subsection{Analysis}
Drawing on Braun and Clarke's six-phase thematic analysis framework \cite{braun2006using}, we collaboratively analysed the interview transcriptions, resulting in an initial set of 237 quotations. These were clustered into categories and subsequently grouped into overarching themes to provide relevant insights. Over the course of three iterative analysis cycles, we refined the dataset to 182 quotations, structured into four overarching themes. From these, three design insights were derived. In parallel, we analysed the artefacts provided by participants, which further informed our exploration of the solution space.

\subsection{Results}

The thematic analysis yielded three design insights to inform and guide our subsequent development of a concept for client-designer communication.

\subsubsection*{Insight 1: Uncover and align with client needs through dialogue and artefacts}
Understanding and aligning with client needs is an essential, yet complex, part of successful design projects. Designers often face clients with vague visions, necessitating their active role in clarification and interpretation. As one graphic designer put it: ``\textit{Some [clients] are really, really good at talking and share a lot in great detail. Others are more like, `We make steel, just make something cool. You have complete creative freedom. But you don't. But you do.'}'' (Graphic Designer 2). This ambiguity can cause misalignment if not actively addressed: ``\textit{If we misunderstand the client, we misspend 100 hours.}'' (Architect 3).

Designers navigate this by employing a combination of strategic questioning, interpersonal sensitivity, and collaborative tools. Several emphasised the need to ``\textit{ask the right questions}'' (Graphic Designer 1) and ``\textit{understand people}'' (Architect 3), often through direct conversation and co-reflection. This is not only to gather information, but also to build trust and surface latent needs: ``\textit{It's really about getting their dreams out so they can be turned into something more tangible.}'' (Architect 3).

Artefacts---such as mood boards and design briefs---play a key role in this process, serving to clarify intentions and facilitate discussion. As one designer explained, ``\textit{I create the brief not just for myself to get an overview of the project, but also to ask the client: `Do you agree that this is the project we are working on?'}'' (Industrial Designer 3). Designers also described how AI can be useful in this context. One participant shared how they used AI-generated visuals to communicate early concepts: ``\textit{I create mood boards where I use AI to describe an atmosphere [...] it is similar to showing early ideas. Sometimes clients struggle to abstract from what they see, but it's really just to facilitate discussion.}'' (Industrial Designer 3). Importantly, AI outputs were framed as sources of inspiration, not final solutions: ``\textit{We use it for inspiration, but we'll never take one of these AI-generated designs and tell the client: `Here it is.'}'' (Industrial Designer 4).

To support early-stage collaboration and facilitate a shared understanding, an elicitation tool should thus include features that prompt reflection through structured questions and co-created artefacts.

\subsubsection*{Insight 2: Utilise the adaptability of AI to support personalised client collaboration}

While certain recurring information is needed in all projects (e.g., budget, timeline), designers emphasised the importance of maintaining a dynamic and adaptable approach to early-stage client collaboration. As one put it: ``\textit{A questionnaire would not work if it asked exactly the same questions every time, because what it is really about is having a dialogue.}'' (Industrial Designer 3). A potential solution should therefore support both standard project elements and the unique needs of each client.

To navigate this tension, designers expressed interest in using AI---particularly to pose tailored questions and help prepare clients for more productive consultations. One designer imagined AI functioning as a `birth assistant' during the ideation phase---structuring information, suggesting questions, and analysing client responses to support expectation alignment and needs articulation. Another described AI as a potential problem-formulation tool: ``\textit{A way to extract the client's wishes or specifications if they are unable to write a design brief. [AI] could analyse the language they use and guide them toward more concrete formulations until they refine their thoughts and reach agreement.}'' (Industrial Designer 2).

However, designers also expressed caution. One architect warned against overcomplication: ``\textit{If it becomes something that can do everything, then I might as well just have a one-hour meeting with the client and take notes myself.}'' (Architect 3).

\subsubsection*{Insight 3: AI should complement, not replace client-designer interaction}
Designers stressed that AI should never replace the essential human interaction between designer and client---emphasising that emotional understanding, nuance, and interpretation are key to the design process.

A primary concern was that understanding a client's needs often comes from unspoken aspects of their responses and motivations. As one designer noted, ``\textit{One thing is what we say, another thing is what we do not say.}'' (Industrial Designer 4), highlighting the limitations of AI in capturing these nuances.

Designers also pointed out that verbal communication is more nuanced and trustworthy than written words. One graphic designer explained, ``\textit{They probably would not answer in the same way [to an AI] as if it were us asking the questions, and they could feel us.}'' (Graphic Designer 2).

Ultimately, while AI can assist in structuring information and guiding initial conversations, the core of the design process remains rooted in human connection, intuition, and interpretation. As one designer put it, ``\textit{We have to keep the human element. That is where a product's edge comes from.}” (Industrial Designer 4).

\begin{table*}[]
\begin{tabular}{ll}
\toprule
Focus area            & \textbf{Insight}                                                            \\ \midrule
Client-need elicitation & \textbf{1)} Uncover and align with client needs through \textbf{dialogue} and \textbf{artefacts}          \\
Context-aware mediation & \textbf{2)} Utilise the adaptability of AI to \textbf{support personalised} client collaboration \\
Human-in-the-loop collaboration & \textbf{3)} AI should \textbf{complement, not replace} client-designer interaction               \\ \bottomrule
\end{tabular}%
\vspace{1em}
\caption{Qualitative insights from phase one of the conducted study.}
\label{tab:phase_1_insights}
\end{table*}

Overall, these insights, as listed in Table \ref{tab:phase_1_insights}, indicate broad agreement among designers that while AI can enhance early-stage collaboration---by prompting reflection and tailoring digital interactions---its true value lies not in replacing human judgement or interpersonal connection, but in scaffolding communication and co-creation. By equipping clients to articulate their needs, preferences, and constraints more clearly, the integration of AI might reduce ambiguity, prevent miscommunication, and foster stronger client-designer alignment from the outset. To avoid diminishing human creativity and agency, such tools must remain flexible and transparent in their role, avoiding over-automation or one-size-fits-all solutions.

\section{Phase 2: Evaluating systems for client preparation and information elicitation}
Building on insights from the phase one interviews, we developed an application to explore how information elicitation tools can be designed to support early-stage client-designer collaboration. From \textbf{Insight 1} on \textit{client need elicitation}, we designed the system to support digital versions of artefacts such as brand cards and mood boards, facilitating need articulation and reflection through choice-based interaction. Based on \textbf{Insight 2} we designed the system in a conversational format, utilising a large language model for \textit{context-aware mediation}. With \textbf{Insight 3} stressing the importance of \textit{human-in-the-loop collaboration}, the system was designed to help clients reflect on and express their needs before a meeting---facilitating effective communication rather than replacing human-to-human interaction. Synthesising these insights, the system took the form of a questionnaire-style conversational agent offering both free-text, choice-based, and visual responses, inquiring clients about their organisation in an effort to motivate project reflection.

\begin{figure}[!ht]
\includegraphics[width=0.6\linewidth]{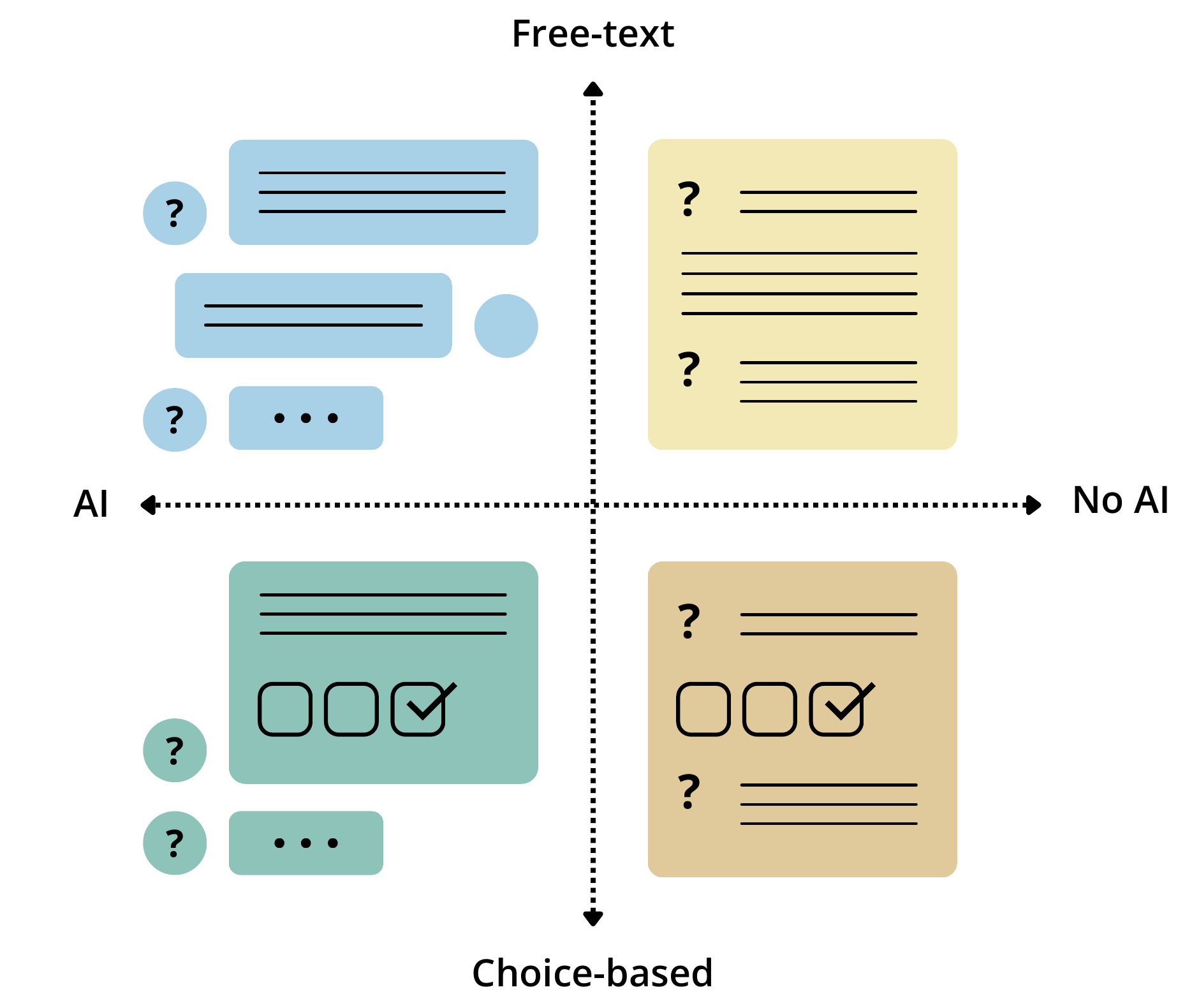}
\caption{Illustrations of the four experimental conditions in the 2x2 factorial design.}
\label{fig:study2}
\end{figure}

Rather than evaluating the system as a whole, we found it more insightful to isolate and examine the impact of key features: (1) the use of interactive, choice-based modules, and (2) the integration of AI through a conversational agent. This approach allowed us to understand the contribution or limitation of each feature independently. To this end, we employed a 2x2 factorial design to examine the effects of \textbf{response type} (free-text versus choice-based) and \textbf{AI presence} (AI versus no AI) on participants' user experience and feelings of preparedness. These experimental conditions are illustrated in Figure \ref{fig:study2}.

In exploring these features, we focused on two key measures: \textbf{user experience} and \textbf{client preparedness}. The first was used to assess whether clients found value in engaging with a digital system for information elicitation; the second to consider whether such engagement could meaningfully support early-stage client-designer collaboration.

\subsection{Designing the application}
\begin{figure*}[!ht]
    \centering
    \noindent\makebox[\textwidth]{%
        \begin{minipage}{0.48\textwidth}
            \centering
            \subfloat[Free-text (AI)]{%
                \includegraphics[width=1\linewidth]{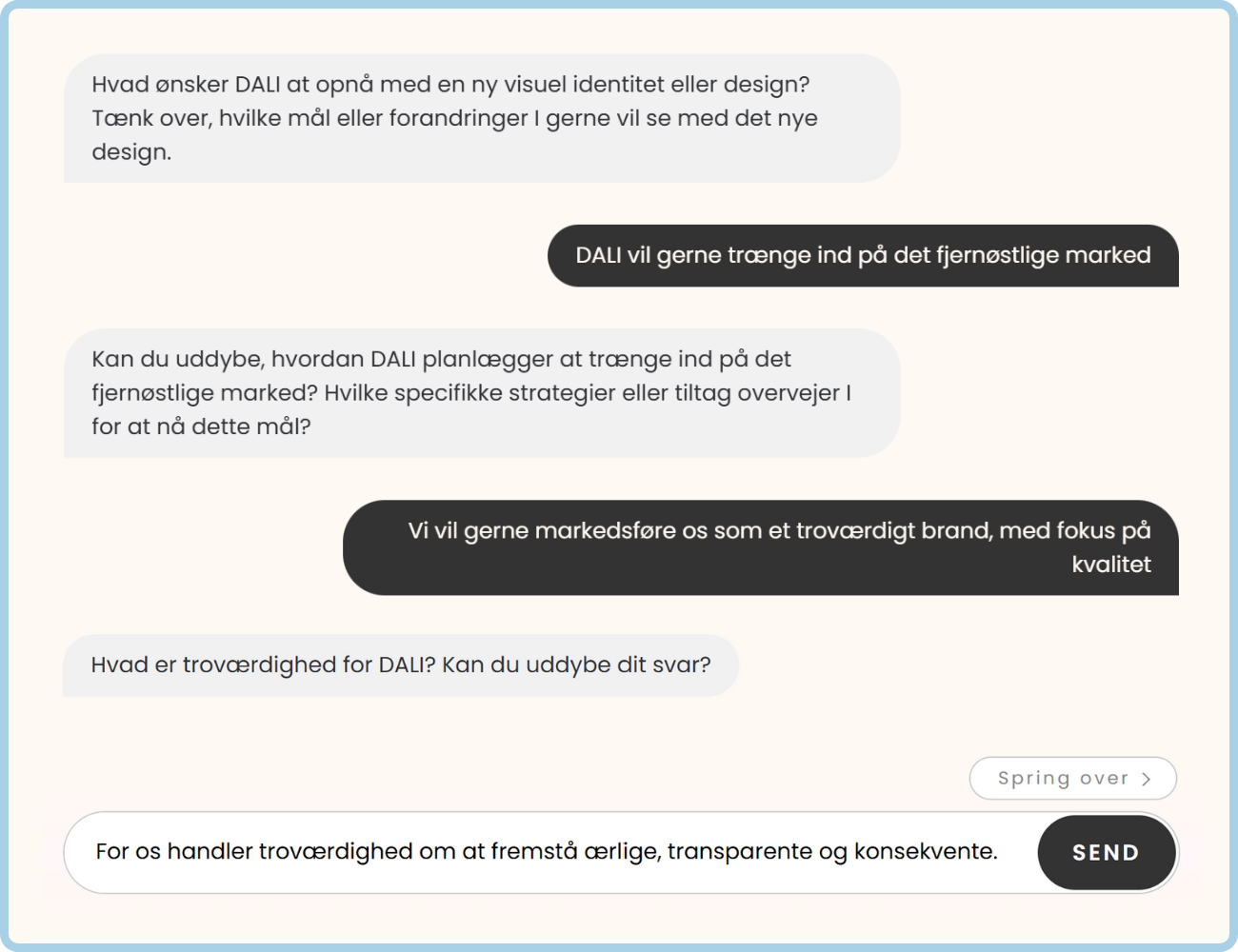}%
                \label{collectionPage}
            }
        \end{minipage}%
        \hfill
        \begin{minipage}{0.48\textwidth}
            \centering
            \subfloat[Free-text (No AI)]{%
                \includegraphics[width=1\linewidth]{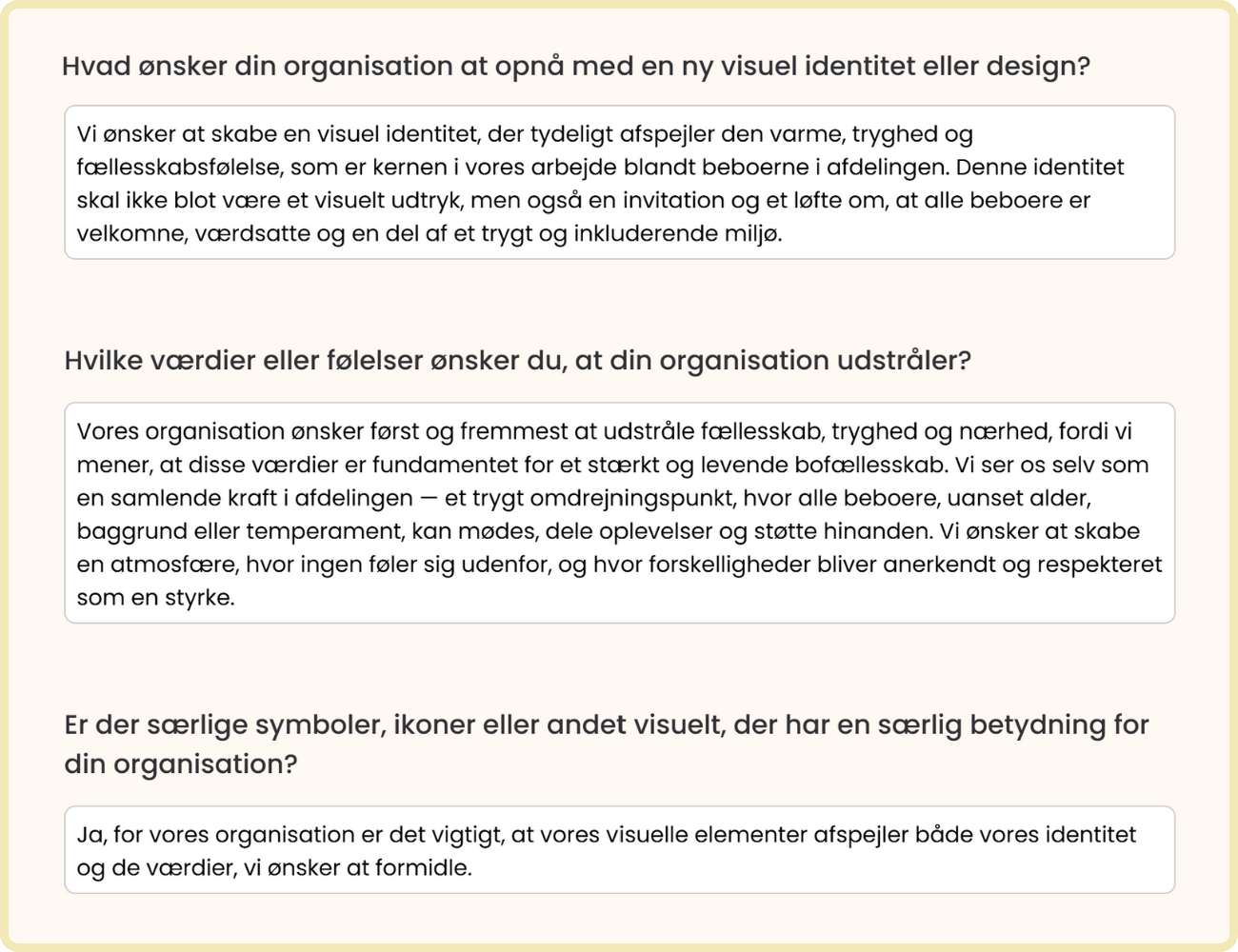}%
                \label{chatPage}
            }
        \end{minipage}
    }

    \vspace{1em}
    
    \noindent\makebox[\textwidth]{%
        \begin{minipage}{0.48\textwidth}
            \centering
            \subfloat[Choice-based (AI)]{%
                \includegraphics[width=1\linewidth]{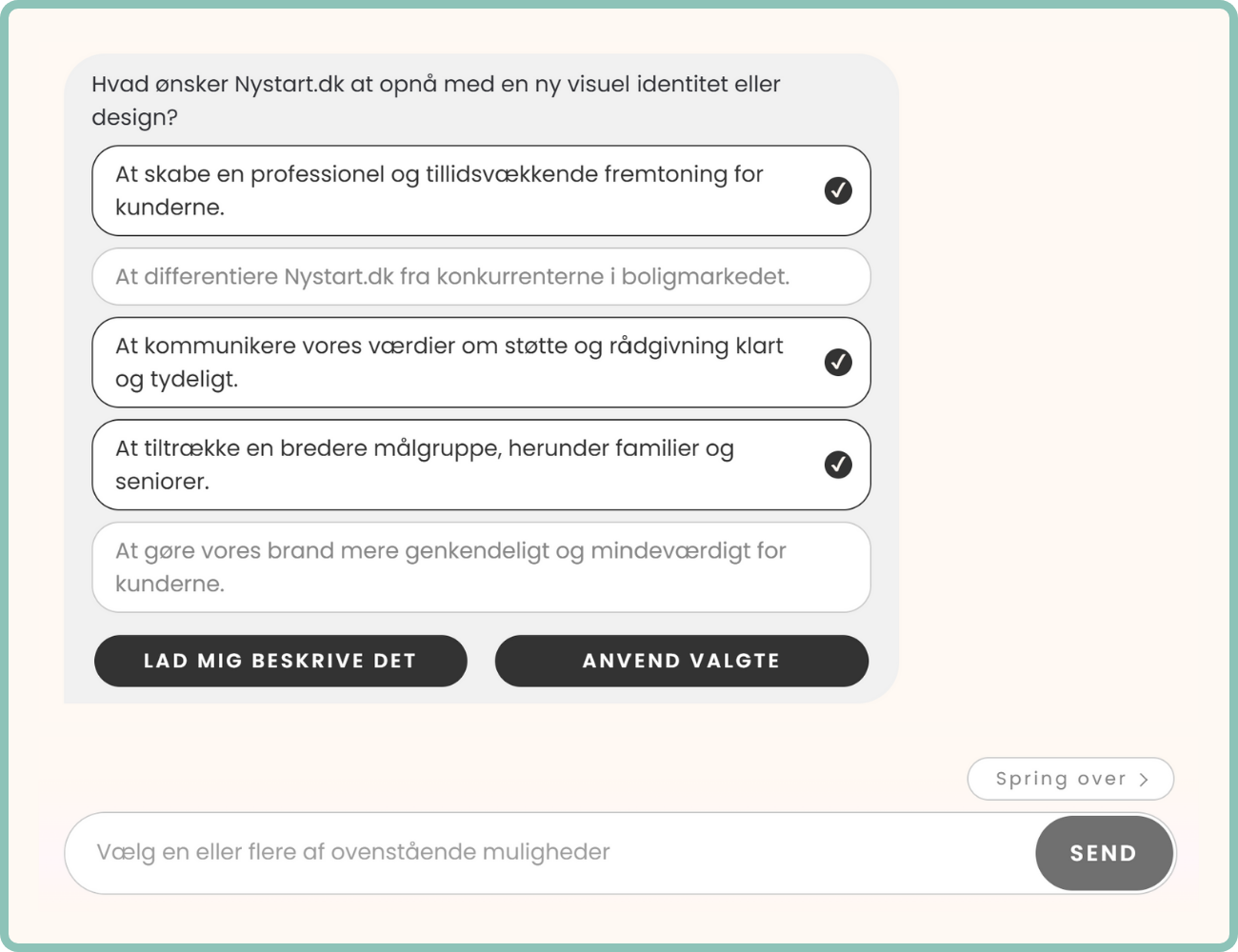}%
                \label{factPage}
            }
        \end{minipage}%
        \hfill
        \begin{minipage}{0.48\textwidth}
            \centering
            \subfloat[Choice-based (No AI)]{%
                \includegraphics[width=1\linewidth]{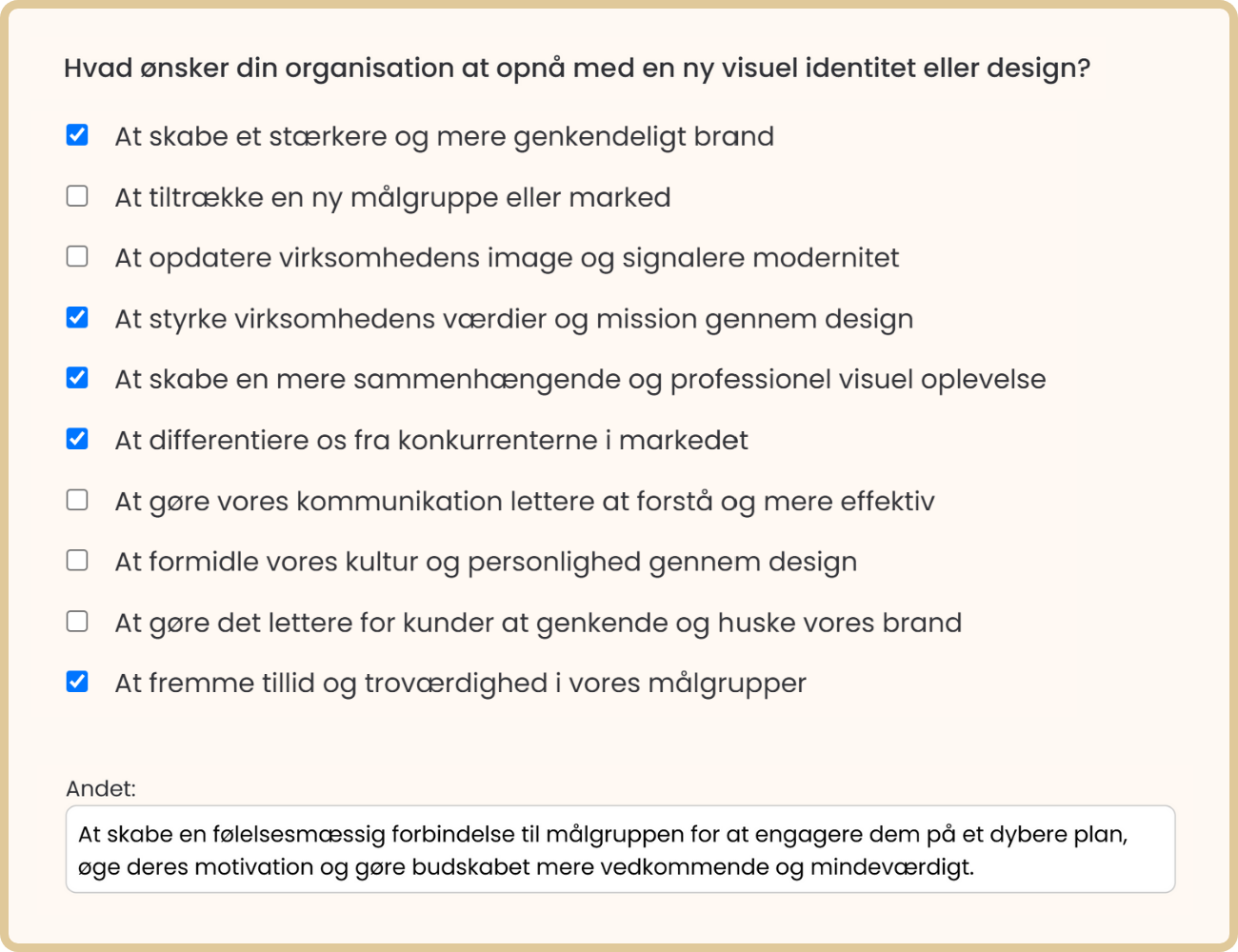}%
                \label{mapPage}
            }
        \end{minipage}
    }
    \Description{The four versions of the developed system.}%
    \caption{Screenshots of the system asking the same question in each of the four experimental conditions.}
    \label{fig:system_versions}
\end{figure*}

To examine the effects of AI presence and response type on UX and preparedness, we developed four versions of an elicitation tool using the \textit{Nuxt}\footnote{\url{https://nuxt.com/}} framework for \textit{Vue.js}\footnote{\url{https://vuejs.org/}}. In practice, we envisioned a system involving two user groups, with designers defining questions for clients to answer. However, to narrow the scope of this study, we focused solely on developing the client-side of the application. The four versions were designed to be visually similar, differing only in regard to the experimental variables---response type and AI presence---as shown in Figure \ref{fig:system_versions}. As can be seen in Figure \ref{fig:teaser_figure}, a sidebar progress bar is present in each version, providing users with a clear indication of their completion status.

\captionsetup[subfloat]{labelformat=empty}

\subsubsection{Questions and topics}
In practice, we envisioned the topics covered by the system to be defined by designers themselves, as their individual needs and wants vary. As our focus was on the client-side, however, we instead curated a list of 17 predefined questions based on the conducted designer interviews. While certain core topics (e.g., needs, budget, and deadlines) consistently recur across design projects, others are necessarily domain-specific. For these, we selected questions relevant to the field of graphic design, drawing on materials shared by designers as well as client-designer collaboration platforms such as HolaBrief\footnote{https://www.holabrief.com/}. The same set of questions was used across all four versions of the system, and participants were free to skip any they wished. Based on previous research, we chose an amount such that completion would take around 20 minutes, as surveys longer than this risks negatively impacting both response rates and response quality \cite{galesic2009effects, revilla2017ideal}.

\subsubsection{Version differences}
The AI versions of the system were designed to resemble state-of-the-art LLMs like \textit{ChatGPT} and \textit{Gemini}. In these versions, questions are asked in a conversational format and users must progress by engaging with the virtual agent. The \textit{no AI} versions were instead modelled after conventional online questionnaires.

Inspired by designer artefacts, choice-based versions supplement free-text questions with four additional question types, including:

\begin{itemize}
    \item \textbf{Multiple choice questions}, where users select from a list of text-based responses.
    \item \textbf{Brand card questions}, where users choose between two polar-opposite adjectives.
    \item \textbf{Mood board questions}, where users select from a set of images.
    \item \textbf{Colour selection questions}, where users pick from a palette of colours.
\end{itemize}

For the \textit{choice-based (no AI)} condition, these options are all predefined. For the \textit{choice-based (AI)} condition, they are \textit{smart options}, generated dynamically based on users' previous responses. In both conditions, users are free to provide free-text responses instead, in case none of the listed options are suitable.

\subsubsection{Prompt engineering}
For the AI-present versions of the system---powered by \textit{gpt-4o-mini} through the \textit{OpenAI API}\footnote{https://openai.com/api/}---we prompted the conversational agent to act as a helpful design assistant, guiding users in preparation for their first meeting with a graphical designer. In addition to defining the agent's overall tone and behaviour, the prompt used in the choice-based condition includes detailed instructions on how the LLM should handle different question types. When asking brand-card questions, for example, the agent is instructed to suggest two polar-opposite words tailored to the user's previous responses. For mood board questions, it is instead prompted to generate an appropriate search string---which in turn is used to retrieve images via the \textit{Unsplash Image API}\footnote{https://unsplash.com/developers}.

To manage the number of follow-up questions and reduce completion time, we configured the system to only ask follow-ups when a user's cumulative response to a given question contains fewer than 60 characters. This threshold was chosen to balance efficiency and response depth, and applied only to free-text questions, excluding those regarding budget and deadline. If a response exceeds this limit, the LLM is instructed to proceed to the next question; otherwise, it is prompted to probe further into the user's answers. 

\subsubsection{Output}
Once users have answered all questions, their responses are presented to them in four different text-based formats. These are shown in the following order:

\begin{itemize}
    \item \textbf{Raw data}, i.e., a verbatim record of the user's responses.
    \item \textbf{Summarised data}, i.e., an AI-generated summary of the user's responses to each topic.
    \item \textbf{Interpreted data}, i.e., an AI-generated synthesis and interpretation of the user's input, highlighting underlying patterns or themes.
    \item \textbf{Actionable data}, i.e., an AI-generated list of reflective questions generated to help the user prepare for their upcoming meeting.
\end{itemize} 

Because many designers work with design briefs or similar documents to help create a shared understanding between them and their clients, we included these outputs to explore how AI might assist designers in making the most of client-provided data.

\subsection{Participants}
Employing a between-subject design, we evaluated one system version per participant. 50 participants successfully completed system evaluation and answered the appurtenant post-evaluation questionnaire\footnote{15 for \textit{free-text (AI)}, 12 for \textit{free-text (no AI)}, 12 for \textit{choice-based (AI)}, and 11 for \textit{choice-based (no AI)}}. An additional six participants completed the evaluation\footnote{2 for \textit{choice-based (AI)} and 4 for \textit{choice-based (no AI)}} but not the questionnaire; their data is used only in the analysis of response quality. To be eligible for participation, users had to be at least 18 years old and fluent in Danish; the language used by the system. Recruitment was conducted via social media and through networks of colleagues and fellow students. Personal invitations were also extended to all design companies involved in the initial interviews, along with a number of start-ups affiliated with Aalborg University's innovation programme.

Of the 50 participants, 27 identified as male and 23 as female. 22 were students, 20 were company employees, 5 were self-employed, and 3 were unemployed or in other circumstances. 9 were aged 18-24, 27 aged 25-34, 9 aged 35-44, and 5 aged 55 or older. Self-reported AI aptitude averaged 4.26 ($\text{median} = 5$) on a scale from 0 (none) to 7 (expert).

\subsection{Procedure}
Participants accessed the system via a public link; the version being programmatically assigned to ensure an even distribution among conditions. The link remained active for one week, allowing participants to access the system from their personal computers at any time and from any location. Upon opening the link, participants were presented with a scenario and instructions on how to proceed. Specifically, they were asked to imagine themselves as representatives of their workplace, an affiliated association, or a fictional project. Their task was to use the system to prepare for a hypothetical meeting with a professional designer regarding their visual identity. Participants were asked to respond in their own words, encouraging them to process and articulate their thoughts freely. They were given no time limit on the use of the system, but the session would end automatically once all questions had been answered. Having gone through the four output variants, participants were redirected to the post-evaluation questionnaire on their experiences.

\subsection{Measures}
\subsubsection{User experience}
To quantify user experience across the four conditions, we employed the User Experience Questionnaire (UEQ) \cite{laugwitz2008construction}, which captures both pragmatic and hedonic dimensions of user experience. Additionally, we collected qualitative data through open-ended questions on participants' user experience and their thoughts on the system output. All questions were integrated into the post-evaluation questionnaire, which also collected demographic information.

\subsubsection{Client preparedness}
To evaluate participants' perceived sense of preparedness and the value of the information provided, we included Likert-scale items alongside optional open-ended questions to elicit deeper qualitative insights. Both system and user prompts, as well as completion times, were logged in a \textit{Supabase}\footnote{\url{https://supabase.com/}} database.

We employed Gricean Maxims to quantify the \textit{specificity}, \textit{relevance}, and \textit{clarity} of participant responses, adapting the work of Xiao et al. \cite{xiao2020chatbotSurvey} and Jacobsen et al. \cite{jacobsen2025chatbots}. Introduced by H.P. Grice, the Gricean Maxims are a set of communication principles to guide effective communication between sender and respondent \cite{grice1975logic, dybkjaer1996grice}. Because the maxims are difficult to quantify directly, we follow Xiao et al. \cite{xiao2020chatbotSurvey} and create proxies for each---i.e., quality metrics, as can be seen in Table \ref{tab:grice_overview}.

\begin{table*}
\begin{tabular}{lllp{4cm}}
\toprule
\textbf{Maxim}     & \textbf{Definition} & \textbf{Quality metric} & \textbf{Definition}                                             \\ \midrule
\textbf{Quantity}  &  One should be as informative as possible & Specificity           & A response should be as detailed as possible      \\
\textbf{Relevance} & One should provide relevant information & Relevance               & A response should be relevant to a question asked \\
\makecell[l]{\textbf{Manner}\\\phantom{0}}    & \makecell[l]{One should communicate in a clear and\\ orderly manner} & \makecell[l]{Clarity\\\phantom{0}}    & \makecell[l]{A response should be clear\\\phantom{0}}                     \\ \bottomrule
\end{tabular}%
\vspace{1em}
\caption{Gricean Maxims and their quality metrics for analysing conversation quality, as adapted from Jacobsen et al. \cite{jacobsen2025chatbots}.}
\label{tab:grice_overview}
\end{table*}

To quantify the \textbf{specificity} of responses (i.e., their level of detail) we manually rated each response on a three-point scale from 0 (generic descriptions) to 2 (specific concepts with detailed examples).

To quantify their \textbf{relevance} (i.e., the degree to which they address the question) we employed a scale from 0 (irrelevant) to 2 (fully relevant).

Finally, to quantify their \textbf{clarity} (i.e.,  how easy they are to understand) we employed a scale from 0 (illegible text) to 2 (complete sentences with no significant grammatical issues).

Unlike Xiao et al. \cite{xiao2020chatbotSurvey} and Jacobsen et al. \cite{jacobsen2025chatbots}, we did not quantify the \textit{informativeness} of participant responses. In their studies, informativeness was calculated as the sum of each word's surprisal---defined as the inverse of its frequency in the response language. As we were unable to find creditable corpora of Danish words from which to calculate word frequencies, and as translated responses might not accurately represent participant sentiment, we excluded this quality metric. Instead, we chose to measure response length, defined as the number of characters in a response.

All responses were scored by the same researcher\footnote{Scoring examples for each metric are provided in Appendix \ref{apx:grice}}, blind to both the experimental condition and the organisation referenced in each answer. Like Xiao et al. \cite{xiao2020chatbotSurvey}, we rated only a subset of participant responses, selecting 12 of the 17 most important questions for analysis. Specifically, we excluded brand card, mood board, and colour selection questions, as well as the initial question asking for the name of the organisation, and the final open-ended question inviting additional comments.

Of the 56 database entries containing participant responses, six were excluded due to insufficient data\footnote{1 for \textit{free-text (AI)}, 2 for \textit{choice-based (AI)}, and 3 for \textit{choice-based (no AI)}}---either because the data had not been properly logged or because only a small number of questions had been answered. Unanswered questions were excluded from both the quality metric and response length calculations.

\subsection{Analysis} \label{section:analysis}

To investigate differences across the two independent variables (see Figure \ref{fig:ueq-plot}), two-way Type III Analyses of Variance (ANOVA) were conducted for each metric. Type III ANOVA was selected for its advantages when groups have different sizes. The independent variables were \textit{AI presence} (AI vs. no AI) and \textit{response type} (free-text vs. choice-based), in a between-subjects 2x2 factorial design. Table~\ref{tab:anova-ueq} presents the results. For any statistically significant effects ($p < .05$), pairwise comparisons were conducted using estimated marginal means, adjusting p-values using the Holm-Bonferroni method to control for family-wise error. 

While the assumption of normality was violated for two of the six UEQ metrics (\textit{attractiveness} and \textit{perspicuity}), as indicated by Shapiro-Wilk tests ($p < .05$), Levene's test showed homogeneity of variances in all cases, and group sizes were approximately equal. ANOVA has been shown to be robust to moderate deviations from normality under these conditions \cite{blanca2017nonNormalDataANOVA}. Thus, we consider the results reliable despite minor assumption violations. 

To analyse the qualitative data from the questionnaire, we employed Braun and Clarke's thematic analysis framework \cite{braun2006using}. This involved systematically grouping questionnaire responses by area of interest (user experience, preparedness and informational value, and output feedback), followed by a process of coding and subsequent theming to derive relevant insights. 

\subsection{Results}
\subsection*{User Experience}
\begin{figure}
    \centering
    \includegraphics[width=\linewidth]{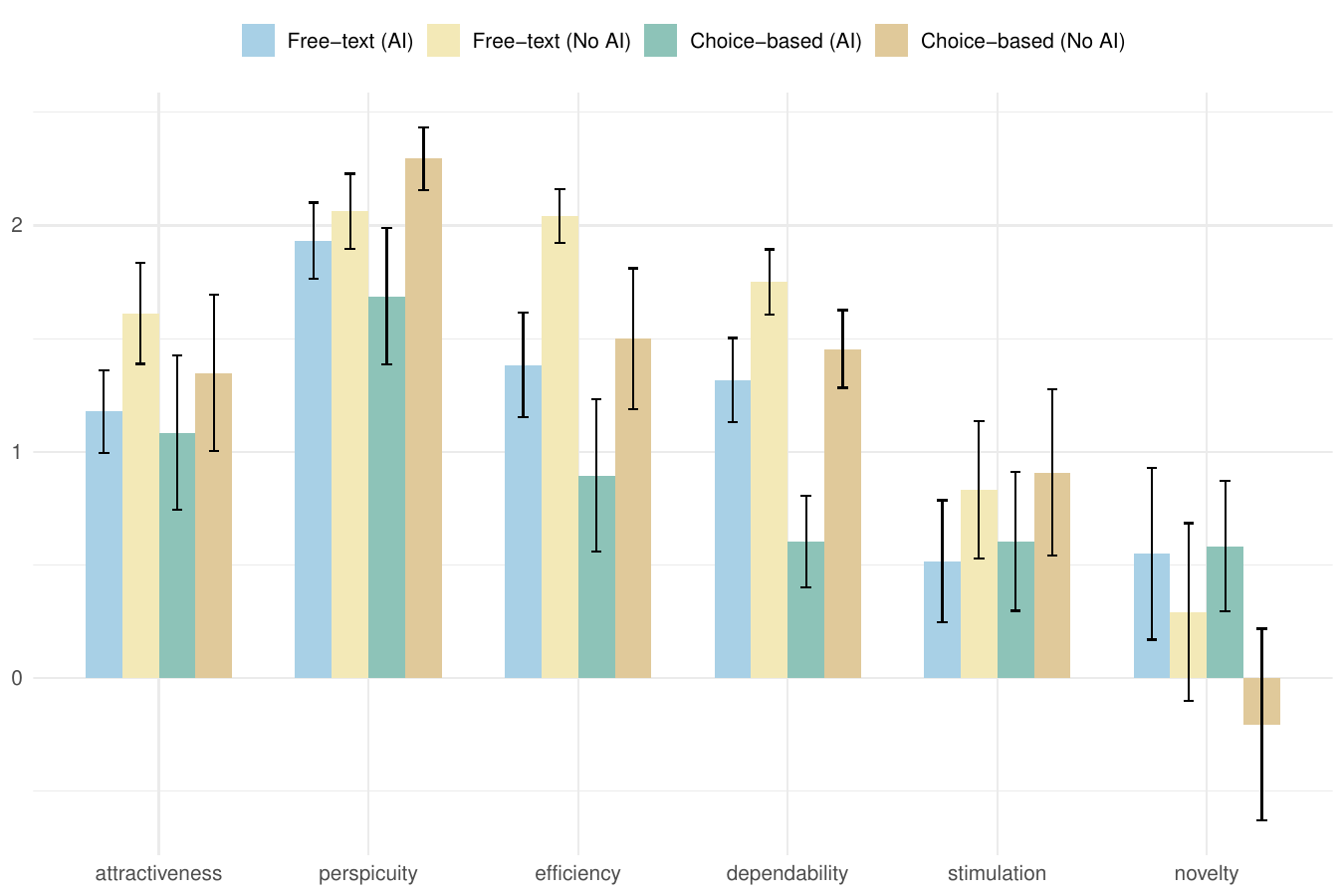}
    \caption{Plot of the UEQ scales for all conditions. Y-axis shows the mean with standard error bars.}
    \label{fig:ueq-plot}
\end{figure}

AI presence, response type, and their interaction effects did not reach statistical significance for attractiveness, perspicuity, stimulation, or novelty. 

For efficiency, a statistically significant main effect was found for AI presence. The subsequent pairwise comparison, using Holm adjustment, revealed a statistically significant difference between \textit{No AI} and \textit{AI} groups. Specifically, efficiency was higher when AI was not present compared to when it was (mean difference = $0.631,\ 95\%\ CI \left[0.107, 1.16\right]$, adjusted $p=0.0194$).
The main effect of response type was not statistically significant and no significant interaction effect was observed. 

For dependability, statistically significant main effects were identified for both AI presence and response type. The post-hoc test for AI presence revealed that the \textit{No AI} condition resulted in significantly higher dependability scores compared to the \textit{AI} condition (mean difference of $0.642,\ 95\%\ CI\ \left[0.276,1.01\right]$, adjusted $p<0.001$). For response type, the post-hoc test indicated that the \textit{free-text} response type led to significantly higher dependability scores than the \textit{choice-based} response type (mean difference of $0.504,\ 95\%\ CI\ \left[0.138,0.87\right]$, adjusted $p=0.008$).

The analysis showed no significant interaction effect between AI presence and response type. Collectively, these results suggest that users perceived the system as more dependable in the absence of conversational AI and choice-based responses---a finding supported by participants' qualitative feedback, which provides deeper insight into how each condition was experienced.

\begin{table}
\begin{tabular}{@{}llll@{}r}
\toprule
\textbf{UEQ Scale} & \textbf{Source}       & \textit{\textbf{F-value}}  & \textit{\textbf{p-value}}     & $\eta_p^2$ \\ \midrule
Attractiveness     & AI presence           & 1.639                      & \phantom{<}.207               & \phantom{<}.030        \\
                   & Response type         & 0.430                      & \phantom{<}.515               & \phantom{<}.009       \\
                   & Interaction effect    & 0.097                      & \phantom{<}.757               & \phantom{<}.002       \\ \midrule
Perspicuity        & AI presence           & 3.254                      & \phantom{<}.078               & \phantom{<}.070        \\
                   & Response type         & 0.001                      & \phantom{<}.975               & <.001      \\
                   & Interaction effect    & 1.373                      & \phantom{<}.247               & \phantom{<}.030        \\ \midrule
Efficiency         & AI presence           & 5.870                      & \phantom{<}.019\textbf{*}      & \phantom{<}.110        \\
                   & Response type         & 3.901                      & \phantom{<}.054               & \phantom{<}.080        \\
                   & Interaction effect    & 0.011                      & \phantom{<}.918               & <.001      \\ \midrule
Dependability      & AI presence           & 12.49                      & <.001\textbf{***}                & \phantom{<}.210        \\
                   & Response type         & 7.700                      & \phantom{<}.008\textbf{**}      & \phantom{<}.140        \\
                   & Interaction effect    & 1.318                      & \phantom{<}.257               & \phantom{<}.030        \\ \midrule
Stimulation        & AI presence           & 1.002                      & \phantom{<}.322               & \phantom{<}.020        \\
                   & Response type         & 0.069                      & \phantom{<}.794               & \phantom{<}.002       \\
                   & Interaction effect    & <0.01                      & \phantom{<}.985               & <.001      \\ \midrule
Novelty            & AI presence           & 1.911                      & \phantom{<}.173               & \phantom{<}.040        \\
                   & Response type         & 0.374                      & \phantom{<}.544               & \phantom{<}.008       \\
                   & Interaction effect    & 0.490                      & \phantom{<}.488               & \phantom{<}.010        \\ \bottomrule
\end{tabular}
\vspace{1em}
\caption{Results of Two-Way ANOVA for UEQ Scales ($df=1, 46$). Significant effects are denoted by * ($p < .05$), ** ($p < .01$) and *** ($p < .001$).}
\label{tab:anova-ueq}
\end{table}

\subsubsection*{Free-text (AI)}
Participants generally found the \textit{free-text (AI)} version of the system useful and easy to use. The conversational format was appreciated as it helped prompt reflection on design needs: ``\textit{I think the program is intuitive to use, and the follow-up questions prompt reflection.}'' (A13)\footnote{Participants are assigned IDs of A for \textit{free-text (AI)}, B for \textit{free-text (no AI)}, C for \textit{choice-based (AI)}, and D for \textit{choice-based (no AI)}}. However, many participants felt the experience became repetitive over time. The pace was described as slow, and the questions often seemed too similar or overly general: ``\textit{I liked it, but at times I felt the pace of the questions was too slow.}'' (A1). ``\textit{Very repetitive questions. It could have picked up on some information earlier, so it didn't feel like such a rigid walkthrough.}'' (A3). One participant also mentioned the lack of navigational control resulting from the chat format: ``\textit{Easy to use, but it needs a `back' button so you can change previous answers. I missed being able to move between tabs when answering subsequent questions.}'' (A12).

\subsubsection*{Free-text (No AI)}

Participants responded positively to the \textit{free-text (No AI)} condition, frequently describing it as straightforward, simple, and easy to use: ``\textit{Easy and straightforward. It was nice to just fill out the different fields without having to think too much, beyond answering the specific questions.}'' (B8). One participant noted that the format could be useful as a conversation starter or a preparatory tool: ``\textit{It was fine. I can see how it could be useful as a kickstarter for a conversation.}'' (B4). However, another participant expressed uncertainty about how to respond to certain questions, which made them difficult to answer:  ``\textit{I was unsure how much freedom I had in my answers, so some questions were hard to answer properly, or I doubted whether I had answered them correctly.}'' (B6). Some also found the experience somewhat flat or impersonal: ``\textit{It was totally fine---though it really just felt like filling out a questionnaire.}'' (B3).

\subsubsection*{Choice-based (AI)}
The version where participants primarily had to select answers from AI generated options received mixed feedback. Several participants appreciated its simplicity and ease of use, highlighting how the smart options and question design made the experience accessible and efficient: ``\textit{It was easy and manageable---really nice that it generated answer options for many of the questions.}'' (C7). Others found the system's step-by-step flow valuable, encouraging reflection and providing a more engaging experience than a static form: ``\textit{It was easy and informative and prompted relevant reflections. It felt more alive than just filling out a template or a `dead' survey.}'' (C9). ``\textit{Overall, it's an exciting and relevant idea with great potential---especially to strengthen preparation before a design meeting. I liked being guided step-by-step, and the questions prompted important considerations.}'' (C11). However, multiple participants also pointed out notable limitations---particularly around flexibility and control during the process: ``\textit{Some questions are quite complex but offer limited response options, while others are simple and don't need much elaboration.}'' (C12). ``\textit{If I answered incorrectly or entered something wrong, I couldn't go back and fix it.}'' (C3). When smart options did not align with participants' intentions, the interaction could feel restrictive or overly mechanical: ``\textit{At times it felt a bit too closed and mechanical, especially when the options didn't quite fit. The `Let me describe it' function was a clear strength and should be emphasised even more. [...] More free-text fields would be a big help. Instead of the `Let me describe it' button, I would prefer if the [input] field was simply unlocked so you could start typing right away, rather than having to wait for the AI---it's a bit slow.}'' (C11).

\vspace{1em}
\subsubsection*{Choice-based (No AI)}
The fourth version, which relied on structured, predefined answer options (without AI-generated content), was also generally well-received. Many participants described the experience as intuitive and inspiring: ``\textit{I think the program worked really well. It inspired me a lot regarding what's relevant to think about when you're pursuing a business or organisational dream. I felt guided and inspired along the way.}'' (D8). However, some participants expressed frustration over limited options and the responsibility placed on clients to contribute substantial input on their own: ``\textit{It's very limited. Normally, in an initial conversation, the consultant helps identify the actual need and explains which products or options are relevant. Here, it requires a lot of prior knowledge and research from the user.}'' (D4). ``\textit{I think the UI was fine. But a lot of the workload is pushed onto the client. [...] It would be easier for the client to just have a conversation with a designer who could then use the program to document the answers.}'' (D10).

\vspace{1em}

\subsubsection*{General feedback}
Beyond feedback tied to specific versions, multiple participants offered insights about the overall concept and user interface. Many appreciated the clarity provided by the progress indicator, which helped them stay oriented during the process: ``\textit{It was great with the bar showing how far I had come. Maybe a bit more of an introduction about the purpose of the answers would be good.}'' (A7). At the same time, some participants expressed a wish for a clearer end-point to the session---especially in the AI-driven versions, where navigation was more limited: ``\textit{It could really use a 'receipt' or finish button for when everything is completed.}'' (B1). A helpful suggestion was the ability to upload or reuse existing company information, reducing the need to manually provide details the system could access directly: ``\textit{I missed the option to upload or add documents and existing materials---for example, if you already have something written about your company that could supplement the answers.}'' (C11).

Overall, participants recognised the system's potential to support early-stage reflection and preparation, but had varying opinions on the ability of conversational AI and choice-based responses to support this.  While AI-generated smart options and conversational formats effectively prompted reflection, many participants expressed a preference for greater control and flexibility---as offered by the \textit{no AI} versions---underscoring the importance of balancing guided interaction with user autonomy.

\subsection*{Preparedness and informational value}
In their ratings of the informational value provided by the system, as well as their feelings of preparedness, participant scores averaged between five and six, with no significant differences between conditions.

Additional statistical tests were performed to evaluate the impact of AI presence and response type on the Gricean Maxim quality metrics (clarity, relevance, specificity) and response length. The ANOVA for clarity revealed a significant interaction effect between AI presence and response type, with post-hoc tests revealing the following: 
\begin{itemize}
    \item When AI was present, choice-based responses were significantly clearer than free-text responses (mean difference of $0.357,\ 95\%\ CI\ \left[0.160, 0.554\right]$, adjusted $p<0.001$).

    \item Comparing choice-based responses across AI presence, \textit{choice-based (AI)} responses were significantly clearer than \textit{choice-based (no AI)} responses (mean difference of $0.502,\ 95\%\ CI \ \left[0.297, 0.706\right]$, adjusted $p<0.001$).

    \item Similarly, \textit{free-text (AI)} responses were significantly clearer than \textit{free-text (no AI)} responses (mean difference of $0.208,\ 95\%\ CI\ \left[0.010, 0.405\right]$, adjusted $p=0.007$).
\end{itemize}

These results suggest that AI presence generally improved clarity, and that this improvement was particularly pronounced for choice-based responses.

Regarding response length, a significant main effect of response type was observed. Post-hoc tests for response type indicated that choice-based responses were significantly longer than free-text responses (mean difference of $63,\ 95\%\ CI\ \allowbreak \left[26, 101\right]$, adjusted $p=0.0014$).

\subsubsection*{Sources of uncertainty and paths to feeling better prepared}

When asked what could make them feel uncertain about meeting with a designer to discuss their visual identity, participants offered a range of reflections. Even though the system is designed for reflection and initial thoughts, not to solicit `perfect' answers, participants often worried whether they had provided enough or the right kind of information: ``\textit{Have I given all the information the designer needs?}'' (C3) and doubted their ability to articulate ideas clearly: ``\textit{I don't have much experience in design---maybe my ideas don't make sense at all.}'' (B5). There was also concern that too much upfront input might limit the designer's creative contribution: ``\textit{I could worry that the creative back-and-forth [...] would disappear}'' (C3). Some participants also emphasised the importance of being able to revisit and revise their input: ``\textit{Let the user loop through the program until they are satisfied with the proposal.}'' (A7). More practical concerns also surfaced, especially around budget and expectations: ``\textit{Do my expectations and budget align?}'' (D9). To better support clients in feeling prepared, multiple participants requested more structure ahead of the meeting: ``\textit{An agenda for the meeting and a short explanation of what the outcome of the meeting will be.}'' (B7). Another theme was the desire for greater transparency about the purpose behind questions: ``\textit{Why do they ask about feelings and behaviour when it's about the visual?}'' (A5).

Finally, some participants suggested the system to include information about designers and their previous work to make the experience feel more personal, reduce uncertainty, and foster trust: ``\textit{Maybe a photo of the designer and references to other material the designer has previously made.}'' (B9).

\subsection*{Output feedback}
Having interacted with their assigned system version, participants were presented with four different output formats---raw data, summarised data, interpreted data, and actionable data---each format identical across experimental conditions. They were asked to describe which format they preferred or found most relevant. Based on their responses, we identified the following takeaways.

\subsubsection*{Value in including multiple output types}

Participants generally appreciated having access to all output formats, highlighting that each served a different purpose. One noted: ``\textit{I thought it was really great that different suggestions were presented for how the data could be displayed. For me, it makes perfect sense to have several different suggestions so you can choose the type of presentation that best fits your company/brand/organisation.}'' (D7).

The layered structure---where users first view the raw data, then a summary or interpretation and finally questions for reflection---was also well received:``\textit{It worked really well to get a summary of your own inputs first---it's familiar because many platforms do that. Then it makes sense to get an interpretation of the entered information, followed by questions that can help you reflect or discuss further.}'' (D8)

\subsubsection*{Reframing responses supports reflection}
Most participants valued the summary and interpreted outputs for rephrasing their input into clearer, more contextual language, aiding self-understanding and reflection.``\textit{I liked being presented with an interpretation of my answers, where additional and different words were used to express some of the thoughts I had.}'' (A1). ``\textit{I preferred the interpretation format---it gave me a clear picture of how my answers were translated into something useful for the designer.}'' (A4).

Some noted that this also allowed them to evaluate their input and even sparked new ideas:
``\textit{It worked well because it showed whether the things between the lines were also understood.}'' (B10). ``\textit{I felt I could be more critical of the information because it was no longer just in my own words.}'' (B6).

\subsubsection*{Editable and verifiable AI output}

While participants found AI summaries and interpretations useful, they also noted how these could sometimes be inaccurate or overly assumptive: ``\textit{Option 3 (interpretation) was fine, but it also made up things that I don't really think applied.}'' (B3). Some suggested adding a feedback function to clarify or adjust interpretations: ``\textit{It would be great to have a feedback function to explain what doesn't fit or to add context.}'' (B8). More commonly, however, participants expressed a desire to manually edit the AI outputs: ``\textit{I liked the interpretation, but I would probably want to edit it myself.}'' (C8).

\subsubsection*{Reflective questions are appreciated, but need work}

Participants appreciated the idea of receiving questions to support further reflection: ``\textit{My favourite was `Questions about the information'---they sparked some good reflections.}'' (C7). ``\textit{The information in tab 4 was clearly the most interesting. Everything else was just summary.}'' (C12). However, many found the questions too generic, repetitive, or disconnected from their input: ``\textit{The questions felt too similar to what I had just answered, so it didn't feel like anyone had really listened to my input.}'' (B6).

\section{Phase 3: Designer feedback}
After evaluating the concept with mock clients, we revisited seven of the ten original design companies to gather their feedback. Designers provided insights on the concept's format, usefulness, improvements, and ideas for integration into practice. 

\subsection{Procedure}
The feedback sessions followed a structured format in which key insights from the interviews were first presented, followed by an explanation of the concept and the experimental setup. To support this explanation, the system was either demonstrated by us or designers were given the opportunity to try it themselves. To illustrate the full range of features, the demonstration primarily drew on the \textit{choice-based (AI)} condition, which showcased elements from all experimental conditions.
Designers were then invited to provide feedback on the concept itself, the proposed output options, and potential directions for further development. The sessions were guided by a semi-structured interview protocol. Six sessions took place on-site while one was conducted online.

\subsection{Analysis}
Data analysis involved transcribing each conversation and collectively going through Braun and Clarke's method of thematic analysis \cite{braun2006using}. From an initial set of 207 quotations, we conducted four iterative rounds of analysis, refining the material to 182 quotations. From these, three themes were identified, informing three design insights.

\subsection{Results}
\subsubsection*{Insight 1: Empower clients to express their needs, not to design solutions}

Designers strongly supported the information-eliciting system for improving early-stage client communication and alignment. It was seen as particularly useful for new or unknown clients, where pre-meeting insight is limited. Designers explained ``\textit{If someone I don't know reaches out to me out of the blue, I could definitely see myself sending them something like this. It would help them prepare for the first meeting. But I wouldn't use it for someone I've worked with before, or if we've already had a first meeting---then it might actually scare them away.}'' (Industrial Designer 2).

While choice-based questions can be efficient, they often require further elaboration to be truly valuable. Regarding single-word descriptors for brand values, one designer noted that ``\textit{one might need a bit more explanation of what is meant by things, or the opportunity to specify it, because if I use it afterwards as a designer, I shouldn't have to do too much interpretation of those answers}'' (Industrial Designer 3).

However, while the system was welcomed as a supportive tool, designers were clear that it should not attempt to make creative decisions or generate solutions. Instead, the system should focus on helping clients reflect on their needs and preferences without locking them into premature solutions. ``\textit{Keep it to a lot of those questions where they have to reflect themselves, and not so much questions where they have to define the solution}'' (Industrial Designer 3).

\subsubsection*{Insight 2: Adapt content and interaction to both client and designer}
The second insight highlights the need for adaptability in the system's content and interaction style to suit both the specific designer's brand and the diverse nature of clients. Designers expressed a desire for personalisation, suggesting that the tool should reflect their company's identity: ``\textit{If it could be made to look like one's company. Made to sound like one's company, that would do a lot of good}'' (Graphic Designer 2). This customisation extends to the questions asked, which should be ``\textit{project-specific for each client}'' (Architect 1). Additionally, it was proposed that designers could enhance informational value by explicitly stating their informational needs: ``\textit{We could say [to the AI]: `Make sure to get sufficient information about what emotional value the product gives'}'' (Graphic Designer 2). 

Designers noted that clients vary in engagement and availability. To accommodate this, the system should personalise interactions based on client preferences---one designer suggested the AI could begin by asking: ``\textit{how should we communicate, [...] how long do you want this to take?}'' (Graphic Designer 2).

Multiple designers also pointed out that clients differ in terms of how established their brand and communication materials are. Well-established companies often have existing documentation---such as branding guides or detailed websites---that could be leveraged to avoid asking redundant questions.  Instead, the system could pre-process available information and present it for verification: ``\textit{It could say: `Hey, I found this information' and then you could give a thumbs up or down.}'' (Graphic Designer 2).

\subsubsection*{Insight 3: AI-generated output is valuable, but must be verifiable}
The third insight revolved around the value of AI-generated output. 
One designer suggested that the order in which the information is shown could influence how it is interpreted: ``\textit{I would probably start with the raw output, because that's what we need to interpret [...] What do \underline{we} think about it, before it gets influenced by what the AI thinks. Then, number two---what does the AI think about it. And number three---what can we then ask.}'' (Graphic Designer 2).

Although valuable, a strong emphasis was placed on the need to critically assess any AI-generated interpretations. The AI's interpretations must be transparent and subject to approval, particularly by the client.: ``\textit{If [the client] has said `that's fine, we'll go with that' [then I would trust it]}'' (Graphic Designer 3). This validation step ensures that the client feels understood and that the AI output aligns with the client's actual intentions. Features like an \textit{approve} button for summaries and interpretations were suggested.

\begin{table*}[]
\begin{tabular}{ll}
\toprule
Focus area & \textbf{Insight $\longrightarrow$ Specification} \\ \midrule
\multirow{2}{*}{Client-need elicitation} & \textbf{1)} Uncover and align with client needs through dialogue and artefacts \\
 & \textbf{$\longrightarrow$} \textbf{Empower clients to express their needs}, not to design solutions \\ \addlinespace\addlinespace
\multirow{2}{*}{Context-aware mediation} & \textbf{2)} Utilise the adaptability of AI to support personalised client collaboration \\
 & \textbf{$\longrightarrow$} \textbf{Adapt} content and interaction \textbf{to both client and designer} \\ \addlinespace\addlinespace
\multirow{2}{*}{Human-in-the-loop collaboration} & \textbf{3)} AI should complement, not replace client-designer interaction \\
 & \textbf{$\longrightarrow$} AI-generated output is \textbf{valuable} if/when \textbf{verifiable} \\ \bottomrule
\end{tabular}%
\vspace{1em}
\caption{Specification of insights from phase one (as can be seen in Table \ref{tab:phase_1_insights}) after conducting phase three.}
\label{tab:phase_3_insights}
\end{table*}

Overall, these insights, as listed in Table \ref{tab:phase_3_insights}, underscore the potential of the information-eliciting system to enhance client-designer communication, particularly in the early stages of a project.

\section{Discussion}
In the beginning, we argued that client-designer alignment is crucial to the success of design projects and that the communication burden and challenges, especially in early stages of design projects, can be potentially addressed with digital and assistive technologies. To this end, we conducted a series of studies to explore how such an assistive system can be designed. With our prototype, we were especially interested in studying the impact of conversational AI and choice-based response formats. We did this, as related work suggest that these emerging techniques and features have the potential to help reduce misunderstandings between designers and clients. 

Overall, for all the design conditions and response formats that were tested, we received positive (and similar) UX ratings from the participants. This shows that independent of tested response format, such as the usage of conversational AI, users rated the different versions of our prototype simply as attractive, easy to get familiar with, effortless to use, and dependable (see Figure \ref{fig:ueq-plot}). In the following, we want to discuss in more detail the few observed differences between the design conditions. Statistical analysis of the UEQ dimensions indicated that both AI presence and choice-based responses were associated with lower scores for the dependability dimension, and that the difference was statistically significant. AI presence also resulted in statistically significant lower scores for the efficiency dimensions. No significant effects of these variables were observed on self-reported levels of preparedness. We draw on qualitative data to contextualise and critically interpret the differences on these specific UX dimensions. 

\subsubsection{The impact of conversational AI and choice-based responses on user experience}
Regarding AI presence, participants noted that the conversational agent felt ``slow'' and ``restrictive'', whereas the AI-absent conditions felt ``simple'' and ``straightforward''. This perceived decrease in efficiency and dependability may stem from the inherently less predictable nature of conversational agents, involving longer idle periods than traditional questionnaires. Consistent with this, the provided response options were sometimes perceived as “less flexible” and “more limited” than free-text input. Such restrictions likely hindered users' ability to fully express their mental models or establish a meaningful alignment with the AI's output, echoing the findings of Robert Frary that free-text responses result in higher internal response validity \cite{frary1985multipleChoice}. As a result, users may have experienced a diminished sense of the AI's ability to accurately understand and guide them.

\subsubsection{The role of client participation in shaping design outcomes} 
While participants valued AI-generated suggestions and interpretations, designers expressed concern that clients may rely too much on these. They worried that clients may accept AI-generated insights uncritically, potentially leading to a skewed or superficial understanding of their own needs. Similar concerns were raised regarding the inclusion of visual stimuli such as images and colours, aligning with previous work that such visualisations can have a significant role in shaping mental models \cite{schnotz2008externalrepresentations}. Given these tensions, the integration of LLMs and choice-based responses into elicitation tools must be carefully calibrated to ensure they enhance---rather than inhibit---the alignment of client-designer mental models.

To avoid design fixation---a cognitive bias where individuals focus narrowly on existing ideas or examples, thereby limiting their capacity to consider alternative solutions \cite{alipour2018fixation, tang2025preandpostai}---some designers suggested omitting any questions tasking clients with providing input on potential solutions. While previous work highlights designers' role not only as providers of requirements, but also of opinions and preferences \cite{casakin2017sharedness}, concerns were expressed that involving clients directly in solution design could lead to unnecessary friction. Given growing concerns about job displacement in the design industry due to AI \cite{li2024aijobreplacement}, designers have a clear stake in preserving the relevance and authority of their expertise. This potential conflict of interest may partly explain a broader resistance to tools perceived as diminishing the designer's central role in the creative process.

Not all designers expressed such concerns, however; others described a more collaborative approach, viewing their role as one of actively supporting clients in reflecting on their vision, goals, and constraints---ultimately informing concrete design decisions. In this view, the value of tools like LLMs lies not in generating design solutions, but in scaffolding client reflection and enabling richer, more nuanced conversations between clients and designers. Based on our interviews, these divergent perspectives seem to align with disciplinary and contextual differences. Designers working in domains where clients are not the end users (i.e., industrial and graphic designers) tended to advocate for more controlled, designer-led elicitation. In contrast, designers whose clients are also the end users (i.e., architects) were more open to co-creation. This variation suggests that a one-size-fits-all system may struggle to meet the diverse expectations and practices found across design disciplines.

\subsubsection{Leveraging conversational AI}
Although previous work highlights how strategic follow-up questions are able to reveal deeper insights into participants' experiences \cite{price2002laddered}, our approach did not effectively support this. Although conversational AI makes it possible to pose contextually appropriate follow-up questions \cite{hu2024designing}, the LLM in the constructed prototype was instructed only to prompt participants to ``elaborate'' on previous responses, resulting in generic follow-ups that rarely referenced earlier answers or requested specific details. Similarly, although a conversational format was employed, the static structure of the questions did not well utilise the dynamic capabilities of the interface. While the format visually resembled a dialogue, the content did not. For instance, even if participants had already provided information about their target age group, the system would still pose the same question later---often without much alteration---making the overall experience feel ``repetitive''.

In general, the LLM did not make much use of participants' previous answer when posing questions. Although it had access to the conversation history---and used it well in the formulation of smart options---the system and prompt design deliberately constrained its ability to deviate from the predefined structure. This rigidity was built in to ensure that all predefined topics were covered, but it ultimately limited the system's ability to leverage the dynamic potential of the conversational agent. This highlights a key tension between the desire for structure (ensuring comprehensive topic coverage) and the desire for dynamism (leveraging the adaptive capabilities of LLMs). To better exploit the strengths of conversational AI, a less structured approach may be preferable. If the system could reliable keep track of user responses, and use this to shape the trajectory of the conversation, it might better facilitate both breadth and depth of exploration, leading to a less mechanical and redundant user experience.

\begin{figure*}
    \centering
    \includegraphics[width=1\linewidth]{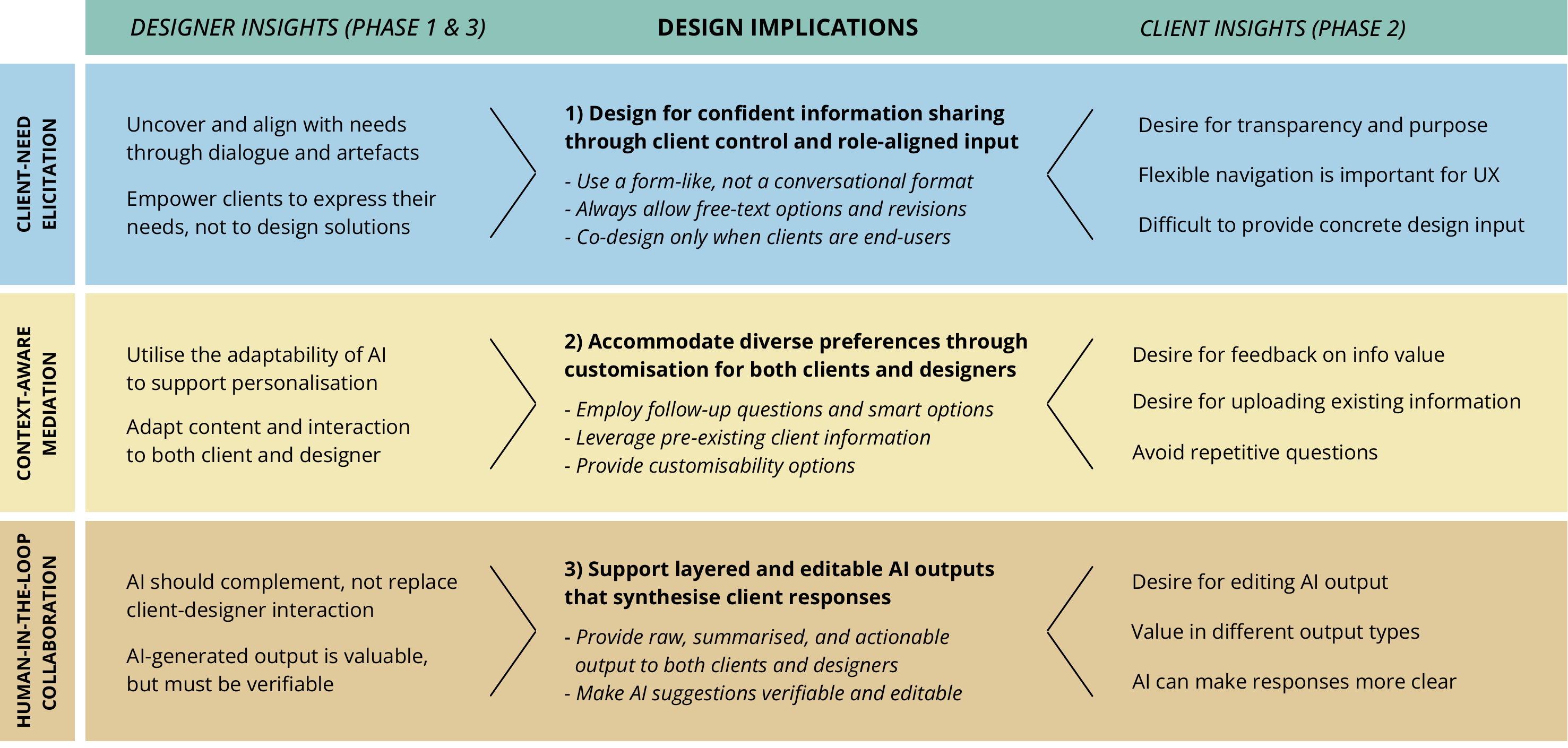}
    \caption{Illustration of how insights---from the phase one designer interviews, the phase two client evaluations, and the phase three designer feedback sessions---informed and shaped the three design implications. Insights also influenced implications across focus areas, but these cross-cutting connections are omitted for clarity.}
    \label{fig:client_designer_implications}
\end{figure*}

\subsection{Design implications}
Based on our study of communication between designers and clients, we identify three key design implications for information elicitation tools that aim to facilitate knowledge transfer and mutual understanding in early stages, as illustrated in Figure \ref{fig:client_designer_implications}.

\subsubsection*{Implication 1: Design for confident information sharing through client control and role-aligned input}
To help clients express their needs clearly and confidently, information elicitation tools should be transparent about their purpose and support flexible navigation, allowing users to revisit and revise responses. Free-text input should always be available to avoid constraining how users articulate their thoughts. Since flexibility can be difficult to achieve in a linear chat format, we recommend embedding an AI assistant in a form-like interface that facilitates reflection and supports the user in providing information through helpful prompts or context-aware smart options. While clients should be encouraged to share diverse information, the type of input should align with their role. When clients are not end-users of the final product, it is more appropriate to focus on objective parameters---such as target group, competitors, and budget---rather than asking for subjective design preferences. This reinforces the designer's expert role while preserving client confidence by avoiding input requests beyond their area of expertise.

\subsubsection*{Implication 2: Accommodate diverse preferences through customisation for both clients and designers}
Because elicitation tools should accommodate the needs of both clients and designers, we recommend utilising the adaptability of conversational AI to customise experiences and thus facilitate client-designer communication. To reflect their professional identity, personality and preferred workflow, designers should have control over visual elements, communicative style, and content organisation---i.e., what information to elicit and to what degree. Likewise, clients should be able to personalise interaction style (e.g., free-text or choice-based input) and time commitment, to ensure the user experience is both meaningful and enjoyable. Incorporating existing materials (such as websites or branding guides) can further minimise redundant input and help customise interactions from the start. 

\subsubsection*{Implication 3: Support layered and editable AI outputs that synthesise client responses}
System outputs should be structured in layers to support diverse cognitive needs and promote interpretability. We recommend a three-stage output format: (1) \textit{raw input}, which ensures transparency and allows independent interpretation; (2) \textit{AI-interpreted summaries}, which synthesises the input and offers complementary insights---these should be editable or confirmable by clients to ensure accuracy and build trust; and (3) \textit{actionable insights}, which propose concrete next steps or considerations. This layered structure encourages both reflection and practical use.

\subsection{Limitations}

While conducting the client and designer evaluations separately provided many practical benefits, this also introduced notable limitations. First and foremost, it meant that no data was gathered regarding the system's effect on the client-designer relationship. Since much of the system's anticipated value lies not in isolated interactions, but in its potential to shape and enhance client-designer cooperation and communication, the separation of the two groups made it difficult for either party to fully assess the benefits of this approach.

Additionally, because the study was not longitudinal, it remains unclear whether introducing a preparatory step at the outset of a design process yields sustained value over time. The impact of such an intervention likely varies depending on project scope and duration---what feels meaningful in a short-term context may prove negligible in longer, more complex projects.

Previous work highlights the importance of self-disclosure in survey research aiming to elicit personal thoughts and feelings \cite{xiao2020chatbotSurvey}, yet our study setup made this difficult. As participants were not real clients of real designers, many responses (such as those regarding the aim, budget, and timeline of the project) were hypothetical by design, making truth-assessment impossible. Although we attempted to simulate real-world use by tailoring scenarios to participants' own contexts, none of the participants were actively seeking a new visual identity during the study. Likewise, the feedback from designers was only based on their own opinion, and they did not test the tool with a client. 
This likely influenced participants' level of engagement and the realism of responses. In a real-world setting---where financial investment and brand strategy are at stake---responses might be more thoughtful, deliberate, and time-consuming.

\subsection{Future work}
One advantage of employing AI in requirements elicitation is the ability to generate dynamic follow-up questions \cite{hu2024designing}, yet deciding which questions to ask---and how many---remains a non-trivial task. In this evaluation, we applied an arbitrary threshold of 60 characters on user responses to decide when to proceed to the next question. This static criterion, however, does not fully leverage the adaptive capabilities of LLMs. A more effective approach might involve basing the decision not on the length of a response, but on its semantic content. Future work could investigate how systems might dynamically determine when to proceed and when to probe deeper, examining whether such adaptive strategies yield more informative or valuable data.

To limit our scope, this evaluation focused on the client-side of the application, pre-defining questions across the board. For evaluations of the systems applicability in a real-world scenario, developing both the client-facing and the designer-facing side of the application would be necessary. We suggest future work to develop such a two-sided system, exploring its longitudinal effects on client-designer collaboration through practical deployment.

\section{Conclusion}

This paper reports on a series of studies to investigate the design of information elicitation tools to support early-stage client-designer collaboration and alignment. We presented a prototype, which received positive UX ratings, and explored differences of alternative and emerging elicitation techniques, including the use of conversational AI to elicit client needs. Drawing on insights from 10 design companies and 50 mock clients, we believe that an embedded AI assistant holds promise for supporting early-stage collaboration. In particular, we highlight its potential to encourage client reflection, enhance the quality of elicited information, and aid designers in synthesising complex inputs. To realise this potential, elicitation tools must offer clients control over the interaction, adapt to diverse user preferences, and provide layered, transparent outputs verifiable by both parties. Through these contributions, we advocate for human-AI collaboration as a means of fostering mutual understanding between clients and designers, strengthening the foundation for meaningful collaboration and design.

\bibliographystyle{ACM-Reference-Format}
\bibliography{bibtex.bib}

\clearpage

\end{document}